\documentclass[aps,amsmath,amssymb,reprint,superscriptaddress,prd]{revtex4-2}

\usepackage{txfonts}
\usepackage{graphicx}
\usepackage{subfigure}
\usepackage{bm}
\usepackage{dcolumn}
\usepackage{braket}
\usepackage{color}
\usepackage[colorlinks,linkcolor=red,anchorcolor=blue,citecolor=blue]{hyperref}

\begin{document}

\preprint{APS/123-QED}

\title{\textbf{Schwinger-Keldysh effective field theory of type-B Goldstone: near-diagonal geometry and Berry term} }%

\author{Pei Zheng}
\email{zhengpei@ucas.ac.cn}
\affiliation{School of Nuclear Science and Technology, University of Chinese Academy of Sciences, Beijing 100049, China}

\author{Mei Huang}
\email{huangmei@ucas.ac.cn}
\affiliation{School of Nuclear Science and Technology, University of Chinese Academy of Sciences, Beijing 100049, China}

\begin{abstract}
In this work, we formulate a finite temperature Schwinger-Keldysh effective field theory for type-B Goldstone modes based on geometric objects globally defined on the coset manifold. In particular, after taking the classical limit, we organize the required two time contour structure by a near-diagonal geometry in which the $r$-type field is interpreted as the physical Goldstone configuration and the $a$-type field is identified as corresponding tangent displacement. Within this geometric viewpoint, the Berry term essential for type-B Goldstones is obtained directly through the transgression of the Berry curvature. Moreover, we show that this exact Berry transgression is compatible with dynamical KMS condition. In order to construct the conservative, dissipative and noise sectors, we classify possible tensors globally defined on the coset manifold. Based on this framework, we discuss in detail two concrete model examples. The dispersion relations of the Goldstone modes and the associated two-point correlation functions are calculated in the presence of dissipation.
\end{abstract}

\maketitle

\section{Introduction}

Symmetry provides one of the most powerful organizing principles of quantum field theory. A phase may nevertheless fail to share the full symmetry of its microscopic dynamics: an order parameter selects one among a family of degenerate states, and a continuous group $G$ is spontaneously broken to a subgroup $H$.  This mechanism underlies phenomena ranging from chiral symmetry breaking to superfluidity and magnetism, and it determines the universal infrared content largely independently of microscopic details. 

The central consequence of spontaneous breaking of a continuous global symmetry is Goldstone's theorem. For internal symmetries in a Lorentz invariant vacuum, every broken generator gives one massless Nambu-Goldstone (NG) boson \cite{Goldstone:1961eq,Goldstone:1962es}.  Their amplitudes vanish at low momentum, so the infrared dynamics is naturally organized as a derivative expansion. The NG fields provide coordinates on the vacuum manifold and transform nonlinearly under full symmetry group $G$, making their effective theory far more constrained than a generic theory of massless scalars. 

The above one-to-one counting is not universal.  When spacetime symmetries are broken, different broken generators can act on the same local fluctuation, and some candidate Goldstones are related by so-called inverse Higgs constraints rather than representing independent modes \cite{Ivanov:1975zq,Low:2001bw,Watanabe:2013iia}. Thus the number of independent physical Goldstone modes may be lesser than the number of broken generators. A familiar setting is one in which matter respects spatial translations and rotations yet singles out a rest frame, thus spontaneously breaking Lorentz boosts. Within this class of systems, fluids, superfluids and solids realize related spacetime breaking patterns, yet they do not generally carry an independent propagating Goldstone for every broken boost \cite{Nicolis:2013lma,Nicolis:2015sra}. This illustrates why the symmetry algebra, the state and locality must all enter the correct counting.

Even for purely internal symmetries, Lorentz non-invariant states provide a second mechanism for modified counting. At finite density, expectation values of commutators of broken charges may pair two broken directions into one canonical degree of freedom. This leads to the modern type-A/type-B classification: type-A Goldstone modes correspond to unpaired broken generators and are generically linear, whereas each type-B Goldstone mode represents a paired set and is generically quadratic \cite{Nielsen:1975hm,Leutwyler:1993iq,Watanabe:2011ec,Watanabe:2012hr,Hidaka:2012ym,Watanabe:2014fva,Brauner:2005di,Brauner:2010wm,Brauner:2024juy,Watanabe:2013uya,Brauner:2014ata}. The ferromagnetic magnon is the canonical type-B example, and related structures occur in more general finite density systems \cite{Nicolis:2012vf,Nicolis:2013sga}.

For isolated systems, several complementary EFT methods are introduced to investigate the properties of Goldstone systems. The old fashioned current algebra and soft theorems determine low-energy amplitudes, while nonlinear sigma models and derivative expansions provide local Lagrangians \cite{Weinberg:1968de,Weinberg:1978kz}. The most general construction is the coset method which is usually called Coleman-Callan-Wess-Zumino prescription \cite{Coleman:1969sm,Callan:1969sn}. This method of construction builds covariant objects from coset $G/H$ and systematically enumerates invariant operators \cite{Nicolis:2013lma}. Further, the Wess-Zumino terms extend this basis when the action, rather than its Lagrangian density, is invariant \cite{Witten:1983tw,DHoker:1994ti,Delacretaz:2014jka,Davighi:2018inx}. For type-B Goldstone modes, such a first-order term supplies the Berry structure that turns two broken coordinates into one phase space pair.

However, real many-body systems are rarely perfectly isolated. Interactions with thermal fluctuations or unobserved degrees of freedom generate attenuation, relaxation and noise, while near a phase transition dissipative transport can control the entire dynamic universality class \cite{Hohenberg:1977ym,Minami:2015nki}. A low-energy description must therefore treat reversible propagation and irreversible relaxation together and must enforce the well-known fluctuation-dissipation relation, rather than adding a damping term phenomenologically after the conservative EFT has been constructed. 

The Schwinger-Keldysh (abbreviated as "SK" hereinafter), or closed-time-path, formalism provides a field theoretic framework suitable for investigating non-equilibrium phenomena \cite{Schwinger:1960qe,Keldysh:1964ud,Martin:1959jp,Chou:1984es,Kamenev:2011,Calzetta_Hu_2023}. In this framework, doubling the fields along forward and backward time branches generates causal real-time correlators and permits an action principle for response, dissipation and fluctuations at the same time. To construct field theory in SK formalism, unitarity becomes a set of structural constraints \cite{Glorioso:2018mmw}. Besides, for a thermal state, the most essential physical constraint, which is called Kubo-Martin-Schwinger relation \cite{Haag:1996,Crossley:2015evo,Glorioso:2017fpd}, is usually implemented locally by dynamical KMS symmetry. Such local symmetry ties dissipative kernels to noise and supplies the EFT form of fluctuation-dissipation \cite{Haehl:2015foa,Crossley:2015evo,Glorioso:2017fpd,Haehl:2016pec,Haehl:2016uah,Gao:2017bqf,Jensen:2018hse}.

Overall, the reach of the SK framework is correspondingly broad. Influence functionals and quantum Brownian motion make it a natural language for open quantum systems \cite{Feynman:1963fq,Caldeira:1982iu,Akamatsu:2014qsa,Pajer:2026fuo,Salcedo:2024nex,Salcedo:2025ezu,Li:2025azq,Li:2026lwl,Kaplanek:2025moq,Kaplanek:2026kpp,Christie:2026dwx,Burrage:2018pyg,Kading:2025cwg}. Keldysh techniques organize transport, disorder and driven many-body dynamics in condensed matter and quantum optics \cite{Kamenev:2009vd,Sieberer:2015svu,Sieberer:2015rtv}. Also, the same contour methods can be used to describe far-from-equilibrium quantum fields, kinetic theory and thermal transport \cite{Berges:2004yj,Jeon:1994if}, fluctuating relativistic hydrodynamics \cite{Kovtun:2012rj,Crossley:2015evo,Glorioso:2017fpd,Glorioso:2018mmw,Glorioso:2020loc,Gao:2018bxz,Donos:2023ibv,Donos:2025jxb}, and even real-time observables in gauge/gravity duality \cite{Son:2002sd,Skenderis:2008dh,deBoer:2018qqm,Baggioli:2023tlc,Bu:2020jfo,Bu:2022esd}. Thus SK is not merely a bookkeeping device for doubling fields. It is a common consistency framework across non-equilibrium quantum physics.

Spontaneously broken phases have increasingly been incorporated into this framework. For example, open system EFTs and the coset construction which has been extended to SK contour identify the doubled symmetry pattern and also provide systematic operator bases \cite{Minami:2015nki,Hongo:2019qhi,Hongo:2018antime,Akyuz:2023nbo,Hongo:2024brb,Landry:2019iel}. However, much of this development is most transparent for type-A Goldstone modes, for which the leading term in action is second order in time derivatives and the Keldysh $a$-type field can be treated locally in the usual way. On the contrary, type-B sectors add a distinct difficulty. Their leading dynamics should be a Berry term, and both its gauge dependence and the nonlinear meaning of the $a$-type field must be controlled globally.

Accordingly, in this work we attempt to address these difficulties  associated with type-B Goldstone sectors through two related constructions. First, we propose the notion of near-diagonal geometry, in which the physical $r$-type field is interpreted as a point on the corresponding manifold, while the $a$-type field is identified as a tangent vector at that point. This viewpoint gives us a covariant nonlinear completion of the usual Keldysh rotation. Most importantly, the desired Berry term is constructed via the transgression method: the globally well-defined Berry curvature is integrated over a chosen region connecting the two SK time contours. In addition, we also use invariant tensors to organize other sectors  in SK effective action. The whole construction is then illustrated by a spin conserving ferromagnet and a dissipative $SU(2)\times U(1)$ linear sigma model.

This paper is organized as follows. In Section \ref{sec:type-B review}, we first provide a brief yet comprehensive review of the essential knowledge concerning type-B Goldstone modes, in order to keep this work self contained. Next in Section \ref{sec:SK construction}, we discuss our formulation of SK effective field theory of type-B Goldstone. Having completed the general discussion of the construction of the SK effective field theory, we apply it to two concrete models in Section \ref{sec:model study}. At last, in Section \ref{sec:summary} we summarize the essential results and conclude with an outlook on future research directions.   

\section{Type-B Goldstone: a brief review}
\label{sec:type-B review}

Assuming the system we are interested in is invariant under a continuous internal symmetry group $G$, with generators denoted by $Q_a$, the corresponding Lie algebra can be written as the standard form,
\begin{equation}
\left[Q_a, Q_b\right]=i f_{a b}^c Q_c .
\end{equation}
If the following criterion is satisfied by some local operator $\mathcal{O}$,
\begin{equation}
\left\langle\left[Q_a, \mathcal{O}\right]\right\rangle \neq 0,
\end{equation}
then the symmetry generated by $Q_a$ is spontaneously broken. In this work, $N_{\rm{BS}}$ is used to denote the number of broken generators.

In a typical relativistic system, we have a one-to-one correspondence between broken generators and Goldstone modes, i.e.,
\begin{equation}
\label{eq:equal number}
N_{\mathrm{NG}}=N_{\mathrm{BS}} .
\end{equation}
However, if we turn our attention to non-relativistic or finite-density systems, Eq. (\ref{eq:equal number}) no longer holds. Besides, if spacetime symmetry, rather than internal symmetry, is spontaneously broken, Eq. (\ref{eq:equal number}) is also invalid \cite{Low:2001bw,Watanabe:2013iia}.  

\subsection{The basic counting rule of Goldstone modes}

As pointed out by previous works \cite{Nielsen:1975hm,Leutwyler:1993iq,Watanabe:2011ec,Watanabe:2012hr,Hidaka:2012ym,Brauner:2005di,Brauner:2010wm,Brauner:2024juy,Watanabe:2013uya,Brauner:2014ata,Watanabe:2014fva,Nicolis:2012vf,Nicolis:2013sga}, to correctly count the number of Goldstone modes, defining the following anti-symmetric matrix is useful, 
\begin{equation}
\label{eq:key matrix}
\rho_{a b}=-i\left\langle\left[Q_a, Q_b\right]\right\rangle.
\end{equation}
In this definition, indices $a,b$ take values only in the space of broken generators.

The correct counting rule of Goldstone modes can be written as follows,
\begin{equation}
N_{\mathrm{NG}}=N_{\mathrm{BS}}-\frac{1}{2} \operatorname{rank} \rho.
\end{equation}
As can be seen, the non-zero rank of matrix $\rho$ will lead the mismatch between $N_{\rm{BS}}$ and $N_{\rm{NG}}$. To explain this counting rule, two different kinds of Goldstone modes are introduced, which will be referred as type-A Goldstone and type-B Goldstone respectively. 

Type-A Goldstones, which appear in relativistic systems, are well known. The most typical example of type-A Goldstone is the pion produced by chiral symmetry breaking. In comparison, type-B Goldstones are less common in realistic systems and have some exotic properties. That is the reason why type-B Goldstones are much more interesting than familiar type-A Goldstone and will serve as the core object of study in this work.  

Type-B Goldstone is closely related to non-zero rank of matrix $\rho$ and the number of such kind of Goldstone is given by
\begin{equation}
\label{eq:number type-B}
N_{\mathrm{B}}=\frac{1}{2} \operatorname{rank} \rho.
\end{equation}
It should be emphasized that Eqs. (\ref{eq:key matrix})-(\ref{eq:number type-B}) apply to the spontaneous breaking of internal symmetries considered in this work. For spontaneously broken spacetime symmetries, additional pairing mechanisms may arise. In particular, as pointed out by a recent work \cite{Glodkowski:2026syy}, a geometric contribution to the Berry curvature induced by the action of broken generators on spacetime translations can pair Goldstone fields even when the conventional matrix $\rho_{ab}$ vanishes. When the discussion is confined to internal symmetries, it follows that the number of type-A Goldstone is 
\begin{equation}
N_{\mathrm{A}}=N_{\mathrm{BS}}-\operatorname{rank} \rho.
\end{equation}
Within the internal symmetry setting considered here, the mismatch between the number of broken generators and independent Goldstone modes originates from the canonical pairing associated with type-B Goldstone modes. The "$1/2$" factor on the right hand side of Eq. (\ref{eq:number type-B}) tells us that approximately two broken generators will give one type-B Goldstone mode. A deeper understanding of type-B Goldstones will become clear in the subsequent discussion.

\subsection{Geometric view of type-B Goldstone}

The physical origin of the central properties of type-B Goldstones lies in the fact that their low-energy dynamics are dominated by topological terms \cite{Watanabe:2014fva}. On the contrary, the low-energy dynamics of type-A Goldstone are purely kinematics. Due to the topological natural of type-B Goldstone, geometric viewpoint is essential and will be investigated in this part.

It is a well known fact that Goldstone modes can be treated as coordinates of coset manifold $G/H$, where $G$ is the full symmetry group and $H$ is the unbroken subgroup. As a consequence, the geometric properties of coset manifold are closely related to the dynamics of Goldstone modes. The defining feature of type-B Goldstone is that the anti-symmetric matrix $\rho_{ab}$ has a non-zero rank. By a real linear transformation such anti-symmetric object can be put into
the following block form:
\begin{equation}
\rho \sim\left(\begin{array}{cccccc}
0 & \lambda_1 & & & & \\
-\lambda_1 & 0 & & & & \\
& & 0 & \lambda_2 & & \\
& & -\lambda_2 & 0 & & \\
& & & & \ddots & \\
& & & & & 0_{\text {zero }}
\end{array}\right) .
\end{equation}
Each non-zero $2\times2$ block shows a symplectic structure. The remaining null directions of $\rho$ just corresponds to type-A Goldstone modes. Intuitively, a natural conjecture is that $\rho_{ab}$ can be related to the Berry curvature of the coset manifold. This is because the commutator of two broken generators measures the difference between moving some object on the coset manifold first one way and then the other, versus the opposite ordering. Therefore, it is not surprising that the commutator is associated with some kind of curvature. 

If we parameterize our state with desired Goldstone excitation as $\left| \pi\right\rangle$, then the standard Berry connection $A(\pi)$ and correspondingly, the Berry curvature $\Omega(\pi)$ can be defined as
\begin{equation}
    A(\pi)=-i\left\langle\pi|\mathrm{d}|\pi  \right\rangle, \quad \text{and} \quad \Omega(\pi)=\mathrm{d}A(\pi).
\end{equation}
The concrete relation between $\rho_{ab}$ and Berry curvature is written as
\begin{equation}
\label{eq:key relation}
    \Omega_{ab}(\pi=0)=\rho_{ab},
\end{equation}
where $\Omega_{ab}$ is the component of Berry curvature $\Omega(\pi)$ of the coset manifold $G/H$. Here, $\pi=0$ refers to the origin of coset manifold which is equivalent to our chosen ground state. The detailed derivation of this essential relation can be found in the Appendix \ref{app:A}. 

Due to such extra structure associated with type-B Goldstone, we can construct new term in Goldstone effective field theory, compared to type-A case. It should be emphasized that the effective field theory here is discussed in the context of general field theory, without involving any construction based on the Schwinger-Keldysh contour. Considering the fact that the low-energy effective field theory is organized by using derivative expansion, the leading time-derivative term can be written as the following form
\begin{equation}
\label{eq:Berry term}
S_{\text {Berry }}=\int \mathrm{d} t A_a(\pi) \dot{\pi}^a =\int_{\gamma} A,
\end{equation}
where $\gamma$ refers to the path traced by coordinates $\pi^a(t)$ moving in the coset manifold and $A=A_a(\pi)\mathrm{d}\pi^a$ is a one-form.

This term is purely topological and is called a Berry term because it has precisely the structure of a Berry phase. To clarify, considering a quantum state $\ket{n(\lambda)}$ which depends adiabatically on parameters $\lambda^a$, then its geometric phase is 
\begin{eqnarray}
&& \gamma_{\text {Berry }}=\int A_a(\lambda) \mathrm{d} \lambda^a, \nonumber\\
&& \quad\text{with}\quad A_a(\lambda)=-i\left\langle n(\lambda) \mid \partial_a n(\lambda)\right\rangle .
\end{eqnarray}
If we just promote the parameters to be dynamical objects, i.e., $\lambda^a=\pi^a(t)$, the geometric phase becomes
\begin{equation}
\gamma_{\text {Berry }}=\int \mathrm{d} t A_a(\pi) \dot{\pi}^a.
\end{equation}
Thus a Berry phase becomes a one-time-derivative term in the effective action, which coincides with Eq. (\ref{eq:Berry term}). 

Near the vacuum, we can further expand the Berry connection 
\begin{equation}
A_a(\pi)=A_a(0)+\partial_b A_a(0) \pi^b+O\left(\pi^2\right) .
\end{equation}
The first term contributes only a total derivative term and thus, can be neglected. After anti-symmetrizing, the first non-trivial term gives
\begin{equation}
\label{eq:topological coupling}
L_{\text {Berry }}^{(2)}=\frac{1}{2} \Omega_{a b}(0) \pi^a \dot{\pi}^b =\dfrac{1}{2}\rho_{ab}\pi^a \dot{\pi}^b,
\end{equation}
with second equality originating from Eq. (\ref{eq:key relation}). 

It can be immediately concluded that such topological term does not contribute to type-A Goldstone because the coefficients corresponding to null directions vanish. This topological term only couple two fields in some certain symplectic block of $\rho_{ab}$. To clarify, for two fields $\pi^1$ and $\pi^2$, suppose $\rho_{12}=-\rho_{21}=B\neq0$. Then
\begin{equation}
L^{(2)}_{\text {Berry }}=\frac{B}{2}\left(\pi^1 \dot{\pi}^2-\pi^2 \dot{\pi}^1\right)=B\pi^1\dot{\pi}^2,
\end{equation}
where second equality holds up to a total derivative. 

Mathematically, Eq. (\ref{eq:topological coupling}) is referred to symplectic potential, which corresponds to the first term of the following familiar form of Lagrangian
\begin{equation}
L=p\dot{q}-H(p,q)=L_{\text{Berry}}-H(p,q).
\end{equation}
Based on this, we see that the two broken coordinates $(\pi^1,\pi^2)$ just describe one canonical pair $(q,p)$ in the phase space, not two independent configuration variables! That's the reason why these two fields, $(\pi^1,\pi^2)$, only correspond to one physical Goldstone excitation, which is classified as type-B Goldstone. In other word, the effective theory which describe type-B Goldstone should be constructed in phase space, rather than configuration space. This is the key property of type-B Goldstone. 

\subsection{Warm-up: ferromagnet as a example}

After discussing some general aspects of type-B Goldstone, a concrete example is desired. In this section, ferromagnet system is investigate which is the simplest type-B Goldstone system. Suppose the full symmetry group is $SU(2)$, the microscopic spin algebra is 
\begin{equation}
\left[S_i, S_j\right]=i \epsilon_{i j k} S_k .
\end{equation}
If the ground state has spontaneous magnetization along the z direction, i.e.,
\begin{equation}
\left\langle S_z\right\rangle=M ,
\end{equation}
then the symmetry breaking pattern is $SU(2)\rightarrow U(1)$, with broken generators $S_x$ and $S_y$. One can easily show that 
\begin{equation}
\rho_{x y}=-i\left\langle\left[S_x, S_y\right]\right\rangle=-i\left\langle i S_z\right\rangle=M .
\end{equation}
Thus
\begin{equation}
\rho=M \mathrm{~d} \pi^x \wedge \mathrm{~d} \pi^y,
\end{equation}
which is nondegenerate on the two-dimensional broken subspace.

In this case, the coset space $SU(2)/U(1)\simeq S^2$, thus the coset element can be parameterized as a unit vector
\begin{equation}
\boldsymbol{n}=(\sin \theta \cos \phi, \sin \theta \sin \phi, \cos \theta) .
\end{equation}
Then the spin Berry term is 
\begin{equation}
S_{\text {Berry }}=M \int \mathrm{~d} t \mathrm{~d}^d x(1-\cos \theta) \dot{\phi}
\end{equation}
with the corresponding Berry connection $A=M(1-\cos \theta)d\phi$. Its Berry curvature is 
\begin{equation}
\Omega=\mathrm{d} A=M \sin \theta \mathrm{~d} \theta \wedge \mathrm{~d} \phi,
\end{equation}
which is $M$ times the area form on the $S^2$.

We further expand $\boldsymbol{n}$ near the north pole, i.e., $\theta$ is treated as small quantity, 
\begin{equation}
\pi^x=\theta \cos \phi, \quad \pi^y=\theta \sin \phi .
\end{equation}
Then the Berry term is written as 
\begin{equation}
L_{\text {Berry }}^{(2)}=\frac{M}{2}\left(\pi^x \dot{\pi}^y-\pi^y \dot{\pi}^x\right) ,
\end{equation}
with Berry curvature $\Omega=M \mathrm{~d} \pi^x \wedge \mathrm{~d} \pi^y$, so the key relation $\Omega_{xy}(0)=M=\rho_{xy}$ is directly verified.

Adding the leading spatial gradient term, we can further investigate the corresponding excitation. Now the Lagrangian is written as
\begin{equation}
L=\frac{M}{2}\left(\pi^x \dot{\pi}^y-\pi^y \dot{\pi}^x\right)-\frac{\rho_s}{2}\left[\left(\nabla \pi^x\right)^2+\left(\nabla \pi^y\right)^2\right] ,
\end{equation}
with $\rho_s$ the so-called spin stiffness. As a result, the corresponding equations of motion can be obtained as
\begin{eqnarray}
&& M \dot{\pi}^y+\rho_s \nabla^2 \pi^x  =0, \nonumber\\
&& {}-M \dot{\pi}^x+\rho_s \nabla^2 \pi^y  =0.
\end{eqnarray}
Defining a complex field, $\psi=\pi^x+i \pi^y$, two equations of motion can be combined to get a single Schr\"odinger-like equation 
\begin{equation}
i M \dot{\psi}=-\rho_s \nabla^2 \psi,
\end{equation}
which gives quadratic dispersion relation. Therefore, the two broken directions $(\pi^x,\pi^y)$ only give one type-B Goldstone mode, which is consistent with the general case presented in the preceding discussion.

\section{The construction of Schwinger-Keldysh effective theory}
\label{sec:SK construction}

In this section, we review some basic ingredients of Schwinger-Keldysh formalism firstly. Based on such fundamental framework, we will discuss the construction of effective field theory suitable for type-B Goldstone system in detail. It should be emphasized that this SK framework is highly comprehensive and can be employed to investigate a wide range of physical problems. Interested readers are encouraged to consult the relevant review articles and seminal works \cite{Chou:1984es,Kamenev:2011,Calzetta_Hu_2023}. 

\subsection{Brief review of Schwinger-Keldysh field theory}

Unlike the S-matrix calculated by particle physicists, we are now interested in studying different aspects of matter properties and the ensemble average of given operators is the underlying observable. Although one could always working in imaginary-time formalism and analytically continuing at the end, working directly in real-time formalism is proved to be practically more useful. 

\begin{figure}[ht]
\centering
\includegraphics[width=\linewidth]{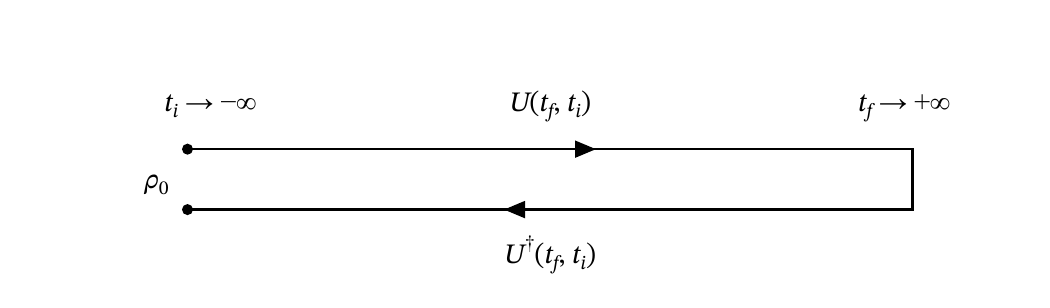}
\caption{A schematic picture of standard Schwinger-Keldysh contour. Arrows show corresponding time evolution directions.}
\label{fig:SK contour}
\end{figure}

If a state is specified by the initial density matrix $\rho_0$, the evolution of such system can be depicted by
\begin{equation}
    \rho(t)=e^{-i H(t-t_i)} \rho_0 e^{i H(t-t_i)},
\end{equation}
with $t_i$ the initial time. The essential structure of such time-dependent density matrix $\rho(t)$ is that evolution operators $U(t,t_i)=e^{-i H(t-t_i)}$ along two opposite time directions appear, which suggests the so-called Schwinger-Keldysh time contour showed in Fig. \ref{fig:SK contour}. A natural consequence of such time contour is the doubling of field contents, i.e., a field variable must be assigned to each of the two branches of the time contour. Accordingly, any operators should be denoted by $\mathcal{O}_i$, with $i=1,2$ the contour index. Besides, we will assume $t_i\rightarrow -\infty$ and adopt this assumption in the subsequent discussion of this work. The case with finite time interval has been discussed in \cite{SciPostPhys.4.1.008}, where a deformed time contour was introduced. In this work, for simplicity, we will not bother ourselves with such possible subtleties.

In order to construct the corresponding generating functional, we need to deform original theory by introducing terms coupling operators of interest to background sources. Thus suitable generating functional is defined as
\begin{equation}
Z\left[J_1, J_2\right] \equiv \operatorname{Tr}\left[\rho_0 \bar{T} e^{-i \int J_2 \mathcal{O}_2} T e^{i \int J_1 \mathcal{O}_1}\right],
\end{equation}
where $T$ and $\bar{T}$ denote time and anti-time ordering respectively. After inserting the complete basis at every time slice, we can convert the generating functional into its path integral representation
\begin{eqnarray}
&& Z\left[J_1, J_2\right]=\int_{\rho_0} \mathcal{D} \varphi_1 \mathcal{D} \varphi_2 \exp \Bigl[i S\left[\varphi_1\right]-i S\left[\varphi_2\right] \nonumber\\
&& {}+i \int \mathrm{~d}^{d+1} x \left(J_1 \mathcal{O}_1-J_2 \mathcal{O}_2\right)\Bigr].
\label{eq:generating functional}
\end{eqnarray}
It should be emphasized that the action $S[\varphi]$ involved in Eq. (\ref{eq:generating functional}) is defined at microscopic level and the path integral gives us a full quantum theory. Formally, any correlator can be calculated by taking functional derivative with respect to external sources. 

When really trying to calculate correlation functions within the SK framework, the following subtlety deserves to be highlighted. The doubling of field contents allows the generating functional to naturally accommodate a variety of distinct correlation functions. Thus the specific type of correlation function under calculation must be unambiguously identified. To make physical structure more explicitly, it is convenient to introduce the so-called Keldysh basis:
\begin{equation}
\mathcal{O}_r \equiv \frac{1}{2}\left(\mathcal{O}_1+\mathcal{O}_2\right), \quad \mathcal{O}_a \equiv \mathcal{O}_1-\mathcal{O}_2.
\end{equation}
Also, the same rotation operation can be applied to the external source field. It is interesting to note that $a$-type external source field should couple to $r$-type operator, rather than $a$-type operator. For example, we can have the following specific two point correlation function written as
\begin{eqnarray}
&& G_{ra}(t_1,t_2)=\dfrac{1}{i}\dfrac{\delta^2 \log Z}{\delta J_a(t_1)\delta J_r(t_2)} \nonumber\\
&& {}=i\braket{\mathcal{O}_r(t_1)\mathcal{O}_a(t_2)}=G_R(t_1,t_2).
\label{eq:retarded correlation}
\end{eqnarray}
The convention for the factor of "$i$" adopted here is consistent with that in the previous work \cite{Crossley:2015evo}. The last equality of Eq. (\ref{eq:retarded correlation}) comes directly from the definition of operators under Keldysh basis. As can be seen, the correlation functions written under Keldysh basis are related to casual correlation functions, which are retarded, advanced and symmetric correlation functions at the level of two point functions. Given that the physical (collective) excitations we are concerned with must be extracted from the pole structure of retarded/advanced correlation functions, formulating the action directly in the Keldysh basis proves to be extremely advantageous. Consequently, the effective theory will be formulated solely in the Keldysh basis in what follows.  

One can directly start with the full quantum functional path integral (\ref{eq:generating functional}), calculate the corresponding nPI effective action and obtain desired evolution equations. Although considerable theoretical progress has been achieved through this top-down approach \cite{Berges:2004yj,Calzetta_Hu_2023}, as a functional method it suffers from significant difficulties in both the analytical derivation and the numerical solution of the corresponding equations, making it inconvenient for applications to certain physical systems. Owing to this, a more simplified bottom-up approach is needed. Fortunately, Hong Liu et al. have developed a systematic bottom-up approach for constructing low-energy effective field theories \cite{Crossley:2015evo,Glorioso:2017fpd}. Schematically speaking, they proposed that the generating functional can be rewritten as
\begin{equation}
    Z[J_1,J_2]\equiv\int \mathcal{D}\phi_1 \mathcal{D}\phi_2 \exp \left\{iS_{\rm{eff}}[\phi_1,\phi_2;J_1,J_2]\right\},
\end{equation}
with $\phi_i$, $i=1,2$ denote gapless dynamical variables only. All  gapped degrees of freedom are assumed to be integrated out and corresponding information are encoded entirely in low-energy effective action $S_{\rm{eff}}$. Given that the Goldstone modes under investigation in this work are gapless modes protected by symmetry, they are particularly well-suited for study within this framework.

The only guiding principle to construct such local effective action is the symmetry. To clarify, we classify the possible symmetries and necessary physical constraints into the following three categories:
\begin{itemize}
    \item \textit{Symmetries possessed by the system under investigation}: this category refers to the symmetries inherent in the system under study, encompassing both gauge symmetries and global symmetries. These symmetries are intrinsic to the system and are independent of whether the SK framework is employed. Once the fundamental symmetries of the system in question are specified, this class of symmetries is correspondingly determined. Symmetries of this type possess no universality and depend entirely on the system itself. 
    \item \textit{Symmetries induced by the initial density matrix}: symmetries of this category are generally emergent in nature, with their origin rooted in the specific choice of the initial density matrix. The most prominent example in this category is the so-called KMS symmetry, which arises from an initial thermal state. A more detailed discussion of this symmetry will be presented in a later section.  
    \item \textit{Constraints arising from the construction based on SK contour}: these constraints are generic within the SK framework. Physically, these constraints originate from fundamental requirements such as unitarity. The following three conditions are proved to be very critical:
    \begin{equation}
    \label{eq:a-field existence}
        S_{\rm{eff}}[\phi_r,\phi_a=0]=0,
    \end{equation}
    \begin{equation}
    \label{eq:imaginary term}
        S_{\rm{eff}}[\phi_r,\phi_a]^{\ast}=-S_{\rm{eff}}[\phi_r,-\phi_a],
    \end{equation}
    \begin{equation}
    \label{eq:convergence of path integral}
        {\rm{Im}} S_{\rm{eff}}[\phi_r,\phi_a]\geq 0.
    \end{equation}
    The first condition just comes from SK contour structure and further implies that $\phi_a=0$ is the solution of the equations of motion derived from varying $S_{\rm{eff}}$ with respect to $\phi_r$. The second condition relies on the fact that effective action $S_{\rm{eff}}$ can be complex in general. In order to guarantee the convergence of path integral, the imaginary part of effective action should be non-negative, which gives us the third condition. 
\end{itemize}
Based on the symmetries and constraints discussed above, a natural way to construct effective action is to write down all possible operators which are permitted by all symmetries and constraints. A necessary truncation is also performed according to the power counting required by the problem under study. In this work, we try to use some geometrical data and basic structures of SK formalism to construct effective theory of type-B Goldstone system, instead of writing down possible operators. 

\subsection{Spontaneous symmetry breaking: near-diagonal geometry}

Let a system has an internal symmetry, which can be described by group $G$. Further, we assume that such internal symmetry is spontaneously broken to some subgroup $H$. It is a well-established fact that that the Goldstone field defines a map $\pi:X\rightarrow \mathcal{M}$ with $X$ spacetime manifold and $\mathcal{M}=G/H$ coset manifold \cite{Coleman:1969sm,Callan:1969sn}. In other word, Goldstone fields can be identified as coordinates of coset manifold. 

Based on above coset formulation, when extending our discussion to SK contour, we choose the relevant manifold to be
\begin{equation}
    \label{eq:coset structure}
    \mathcal{M}\rightarrow \mathcal{M}_1\times \mathcal{M}_2=(G_1/H_1)\times (G_2/H_2).
\end{equation}
Here, a subtle point deserves attention: since the present work is concerned solely with the properties of the Goldstone modes themselves, we directly double the physical coset here. This differs from analyzing the symmetry breaking behavior starting from the full SK generating functional \cite{Akyuz:2023nbo}. A more detailed discussion to clarify the distinction between these two will be made at the end of this section.

Intuitively, the evolution of Goldstone fields along two SK branches can be visualized as two trajectories on the corresponding manifold $\mathcal{M}$. To manifest the geometric meaning of Goldstone fields, we find that using Keldysh basis is convenient. Usually, the $r$-type field is interpreted as classical field or average field which is the desired physical object in low-energy effective theory. Mathematically, this classical field is defined by $\pi_1=\pi_2$. Thus, from our geometric perspective, $r$-type field, $\pi_r$ in this case, define the following diagonal submanifold
\begin{equation}
\label{eq:submanifold}
\Delta_{\mathcal{M}}=\{(p, p) \mid p \in \mathcal{M}\} \subset \mathcal{M}_1 \times \mathcal{M}_2.
\end{equation}
However, Eq. (\ref{eq:a-field existence}) also implies that action defined in this exact diagonal submanifold vanishes. One must work in some larger space in order to write down a non-trivial effective action, which is exactly the $a$-type fields come into play. Inspired by the usual definition of Keldysh basis, $a$-type field describes the deviation of the field defined directly on the contour from the classical $r$-type field. Thus, we identify $a$-type field to be the corresponding tangent vector, i.e., 
\begin{equation}
\pi_a(x) \in T_{\pi_r(x)} \mathcal{M},
\end{equation}
which gives a clear geometrical interpretation of $a$-type field. It should be emphasized here that this identification is not exact in general situation, but it makes sense when dealing with classical limit. Before discussing the concrete relation between classical limit and such identification, we need to introduce a useful exponential map first. 

In general, our coset manifold should be a curved manifold. But the definition of usual Keldysh basis is suitable for flat space and may be treated as linear approximation of general curved case. To go beyond linear approximation, we can define the following exponential map:
\begin{equation}
    \exp _p: T_p \mathcal{M} \rightarrow \mathcal{M}.
\end{equation}
Given a tangent vector $v$, this map will give us another point on the manifold $\mathcal{M}$. To be more specific, the given tangent vector define a geodesic $\gamma_v(\lambda)$ with affine parameter denoted by $\lambda$. Obviously, this geodesic must satisfy 
\begin{equation}
    \gamma_v(0)=p, \quad \dot{\gamma}_v(0)=v .
\end{equation}
The point given by the exponential map is then defined as
\begin{equation}
    \exp _p(v)=\gamma_v(1).
\end{equation}
Applying this exponential map to the present SK formulation and remembering the fact that $\pi_a$ plays a role of tangent vector, we have
\begin{equation}
    \label{eq:exponential parameterization}
    \pi_1=\exp _{\pi_r}\left(\frac{1}{2} \pi_a\right), \quad \pi_2=\exp _{\pi_r}\left(-\frac{1}{2} \pi_a\right).
\end{equation}
Readers may not be familiar with this form but actually similar expressions have already been used when studying non-Abelian hydrodynamics \cite{Glorioso:2020loc,Hongo:2024brb}. Eq. (\ref{eq:exponential parameterization}) will be employed in the subsequent discussion.

To gain some experience, we can consider the simplest case here. Assume coset manifold is flat, say $\mathbb{R}^n$, the geodesic is simply a straight line
\begin{equation}
    \gamma_v(\lambda)=p+\lambda v.
\end{equation}
Consequently, the exponential map simplifies to be
\begin{equation}
    \exp_p(v)=p+v.
\end{equation}
It follows immediately that the standard field representation in the Keldysh basis is recovered:
\begin{equation}
    \label{eq:linear form}
    \pi_1=\pi_r+\dfrac{1}{2}\pi_a, \quad\ \pi_2=\pi_r-\dfrac{1}{2}\pi_a.
\end{equation}
Thus one can treat Eq. (\ref{eq:exponential parameterization}) as curved manifold generalization of Eq. (\ref{eq:linear form}) which naturally includes extra nonlinear contributions. 

After defining above exponential map, we are now ready to discuss relations between classical limit and our geometric identification. In order to elucidate the different geometric roles of the $r$-type and $a$-type fields, we need to investigate how they transform under the symmetry group $G$ operations. Let $K_{\alpha}=K^A_{\alpha}(\pi)\partial_A$ be the fundamental vector fields generated by $G$ on $\mathcal{M}$, with $\alpha$ corresponding to generator of Lie algebra, i.e., $T_{\alpha}\in\mathfrak{g}$. Thus the transformation on each branch of SK contour are
\begin{equation}
\label{eq:transformation rule}
\delta \pi_s^A=\epsilon_s^\alpha K_\alpha^A\left(\pi_s\right), \quad s=1,2 .
\end{equation} 
Next, the classical limit is coming into play. As pointed out previously by \cite{Crossley:2015evo,Glorioso:2017fpd,Glorioso:2018mmw}, the typical fields should scale as $\pi_r\sim \mathcal{O}(1)$ and $\pi_a\sim \mathcal{O}(\hbar)$ in the classical limit. Based on this power counting, we can then expand our fields around classical $r$-type field with corresponding scale implicit. We now write $\pi_1=\pi_r+\pi_a/2+\cdots$ and $\pi_2=\pi_r-\pi_a/2+\cdots$, and also define $\epsilon_r=(\epsilon_1+\epsilon_2)/2$, $\epsilon_a=\epsilon_1-\epsilon_2$. Expanding Eq. (\ref{eq:transformation rule}) gives
\begin{eqnarray}
&& \delta \pi_r^A=\epsilon_r^\alpha K_\alpha^A+\frac{1}{4} \epsilon_a^\alpha \pi_a^B \nabla_B K_\alpha^A+\mathcal{O}\left(\pi_a^2\right), \nonumber\\
&& \delta \pi_a^A=\epsilon_a^\alpha K_\alpha^A+\epsilon_r^\alpha \pi_a^B \nabla_B K_\alpha^A+\mathcal{O}\left(\pi_a^2\right) ,
\end{eqnarray}
which is written covariantly at given point $\pi_r$. According to these two transformation behaviors, one can see the following power counting of transformation parameters holds
\begin{equation}
    \epsilon_r\sim \mathcal{O}(1), \quad \text{and} \quad \epsilon_a\sim \mathcal{O}(\hbar). 
\end{equation} 
Thus, for leading classical order, we can have 
\begin{equation}
\epsilon_a=0: \quad \delta_r \pi_r=\epsilon_r^\alpha K_\alpha, \quad \delta_r \pi_a=\epsilon_r^\alpha\left(\mathrm{d} K_\alpha\right)_{\pi_r} \pi_a.
\end{equation}
This is exactly the push forward law for a tangent vector based at the transformed point $\pi_r$. Thus such transformation behavior of SK Goldstone fields clarify our geometric identifications proposed before. Interestingly, when we take $\epsilon_a=0$, the full symmetry group becomes the diagonal part of original $G_1\times G_2$, which is referred to $G_{\rm{diag}}$. Accordingly, when we confine ourselves to the classical limit, the construction of our effective theory only requires that the basic building blocks be invariant under the group $G_{\rm{diag}}$. This observation simplifies the subsequent derivations a lot.

One may further ask: what does the classical limit correspond to within our geometric perspective? To answer this question, we first assume our manifold has a well-defined metric $g_{AB}(\pi_r)$. For physical systems, this assumption generally holds. Based on the exponential map discussed before, the following quantity,
\begin{equation}
    \left\|\pi_a\right\|_g \equiv \sqrt{g_{A B}\left(\pi_r\right) \pi_a^A \pi_a^B}=d_{M}(\pi_1,\pi_2),
\end{equation}
just gives us the invariant measure of contour separation. Here, $d_M(\pi_1,\pi_2)$ refers to the geodesic distance between two trajectories on the manifold which correspond to two SK contours. Taking into account the power counting behavior of the $a$-type field in the classical limit, we have
\begin{equation}
    \left\|\pi_a\right\|_g\sim\mathcal{O}(\hbar)\ll 1,
\end{equation}
which gives us a quite narrow region near diagonal submanifold defined in Eq. (\ref{eq:submanifold}). Therefore, upon combining the SK formulation with the classical limit, the region of genuine physical interest is restricted to a small neighborhood of the diagonal submanifold. In the subsequent discussion, we refer to this region as the near-diagonal region hereinafter.

Before ending this section, some detailed discussions about Eq. (\ref{eq:coset structure}) are necessary. Actually, if we begin with the full SK generating functional and try to investigate its properties under symmetry breaking, we can find that the results relies on initial density matrix. Confining ourselves to the thermal state, the corresponding symmetry breaking pattern is $G_1\times G_2\rightarrow H_{\rm{diag}}$. Interested readers may consult the original reference for details \cite{Akyuz:2023nbo}. The coset manifold directly given by this symmetry breaking pattern has larger dimension than that given by Eq. (\ref{eq:coset structure}), which can be manifested by the following equations, 
\begin{eqnarray}
    &{\rm{dim}}\left(\dfrac{G_1\times G_2}{H_{\rm{diag}}}\right)=2 {\rm{dim}}( G)-{\rm{dim}}(H),\nonumber\\ &{\rm{dim}}\left(\dfrac{G_1}{H_1}\times\dfrac{G_2}{H_2}\right)=2{\rm{dim}}(G)-2{\rm{dim}}(H).
\end{eqnarray}
Relative to the result furnished directly by the SK generating functional, what Eq. (\ref{eq:coset structure}) lacks is exactly the off-diagonal part of $H_1\times H_2$. Physically, the breaking pattern $G_1\times G_2\rightarrow H_{\rm{diag}}$ characterizes the symmetry realization of the full thermal Schwinger–Keldysh generating functional. Its coset coordinates, however, should not all be interpreted as independent physical Goldstone excitations. In fact, the authors of previous work \cite{Akyuz:2023nbo} have already noticed this subtlety, and they relate the off-diagonal part of $H_1\times H_2$ to the unbroken charge density, rather than interpreting this part as new Goldstone degrees of freedom. Since the present work focuses primarily on the Goldstone degrees of freedom themselves, it is reasonable to neglect the contribution from the off-diagonal part of $H_1\times H_2$. Thus, Eq. (\ref{eq:coset structure}) is a reasonable starting point of our construction. Rigorously speaking, what is studied in this work can be identified as the reduced Goldstone sector of the underlying full theory.

\subsection{Dynamical KMS condition}

In the reduced Goldstone sector EFT considered in this work, conserved density or charge response fields are not retained as independent dynamical variables. There is no need to introduce any external source fields. Owing to this, the correct version of KMS condition should be so-called dynamical KMS (abbreviated as DKMS hereinafter) condition proposed by \cite{Sieberer:2015svu,Glorioso:2017fpd}. Moreover, we find the framework developed by \cite{Haehl:2016pec,Haehl:2016uah}, which will be called HLR hereinafter, to be particularly helpful. Most importantly, the derivation of the HLR is entirely operator-based and does not rely on any particular choice of coordinates. Accordingly, the derivation of DKMS presented in this section will rely on this particular framework.

Based on insights from many previous studies \cite{Crossley:2015evo,Glorioso:2017fpd,Haehl:2016pec,Haehl:2016uah}, DKMS contains two basic ingredients: one is the thermal translation and the other is the anti-unitary operation. The anti-unitarity of the latter operation stems from the necessity of incorporating the time-reversal transformation. For certain systems, such as ferromagnet, spontaneous symmetry breaking gives rise to a ground state that explicitly fails to preserve time-reversal symmetry. In this case, the anti-unitary transformation must include some appropriate discrete symmetry operations other than time-reversal transformation, so that the broken ground state is formally invariant under the resulting generalized time-reversal transformation. For the discussion in this section, it is not necessary to specify the explicit form of the anti-unitary operation. Formally, we define such anti-unitary transformation as $\Theta=(\vartheta,\tau)$. Here, $\vartheta : X \rightarrow X$ denotes the transformation acting on the physical spacetime, while $\tau : M \rightarrow M$ denotes the transformation acting on the coset manifold. 

In general, a timelike vector $\beta^{\mu}$ and a flavor gauge parameter $\Lambda_{\beta}$ are needed to specify a thermal state. $\beta^{\mu}$ vector is related to inverse temperature and $\Lambda_{\beta}$ is the generalization of chemical potential. In HLR, these two quantities are referred to as thermal vector and thermal twist. Further, the thermal vector can generate corresponding diffeomorphism and thermal twist can generate some flavor transformation. Then the full thermal translation operator $\delta_{\beta}$ is defined as
\begin{equation}
    \label{eq:thermal translation}
    \widetilde{O}(t)\equiv e^{-i\delta_{\beta}}O(t)\equiv  e^{\beta (H-\mu_I Q^I)} O(t)e^{-\beta (H-\mu_I Q^I)},
\end{equation}
with $O(t)$ an arbitrary operator. Note that this operator $\delta_{\beta}$ should be regarded as a first-order differential operator, $\delta_{\beta}\sim \mathcal{O}(\partial)$. Thus, the expansion with respect to $\delta_{\beta}$ is equivalent to usual derivative expansion. For convenience, a deviation operator $\Delta_{\beta}$ can be further defined as 
\begin{equation}
    i \Delta_\beta=1-e^{-i \delta_\beta}.
\end{equation}
In the long wavelength limit, such deviation operator can be expanded as 
\begin{equation}
    i \Delta_\beta=i \delta_\beta+\frac{1}{2} \delta_\beta^2+\mathcal{O}\left(\delta_\beta^3\right).
\end{equation}
It should be emphasized that $\delta_{\beta}$ and $\Delta_{\beta}$ agree only at linear order. The distinction between these two needs to be further clarified. $\delta_\beta$ generates the infinitesimal thermal translation, whereas $\Delta_{\beta}$ measures the finite KMS deviation between a field and its thermally translated image which includes contributions from higher-order derivatives.

For the sake of simplicity, we refrain from presenting the detailed algebraic discussions contained in the HLR work \cite{Haehl:2016pec,Haehl:2016uah}. Instead, we just quote their central observation here: the following operator
\begin{equation}
    \label{eq:KMS-shifted}
    O_{a, \beta} \equiv O_1-\widetilde{O}_2=O_1-e^{-i\delta_\beta} O_2=\left(O_1-O_2\right)+i\Delta_{\beta} O_2
\end{equation}
is the proper KMS-shifted operator under a thermal translation. Based on this conclusion, we can proceed to discuss the behavior of the Goldstone system under the operation of thermal translation. 

From Eq. (\ref{eq:thermal translation}), the infinitesimal thermal translation induces the following variation of Goldstone map: 
\begin{equation}
    \delta_\beta \pi^A_r=K_\beta^A[\pi_r] \equiv \beta^\mu D_\mu \pi^A_r+\delta_{\Lambda_{\beta}} \pi_r.
\end{equation}
The first term of this infinitesimal transformation originates from a translation along the direction of the thermal vector, while the second term arises from a possible internal thermal twist. Here $D_{\mu}$ is the coset covariant derivative with respect to appropriate connection. In fact, since the $\pi^A_r$ fields here are themselves scalar coordinates on the coset, the covariant derivative in this case reduces to the ordinary derivative. Nevertheless, for notational consistency, we employ the covariant derivative throughout, which causes no ambiguity. 

Based on Eq. (\ref{eq:exponential parameterization}), $\pi_1$ and $\pi_2$ are two different points defined on curved coset manifold. Thus, their direct subtraction is not geometrically meaningful. As a result, the following inverse exponential map, i.e., logarithmic map, is needed:
\begin{equation}
\log _{\pi_r}: M  \rightarrow T_{\pi_r} M,
\end{equation}
which converts each nearby point into a tangent vector based at the common point $\pi_r$. Therefore, the direct analog of Eq. (\ref{eq:KMS-shifted}) in our Goldstone system is
\begin{equation}
    \pi_{a, \beta}(x) \equiv \log _{\pi_r(x)} \pi_1(x)-\log _{\pi_r(x)} \widetilde{\pi}_2(x),
\end{equation}
which is mathematically meaningful. Since we are concerned with the low-energy effective field theory, our theory is organized in the spirit of the derivative expansion. We therefore treat $\delta_{\beta}$ as a small quantity and perform an expansion of $\widetilde{\pi}_2$ accordingly:
\begin{eqnarray}
&& e^{-i \delta_\beta} \pi_2^A=\pi_2^A-i \delta_\beta \pi_2^A+\mathcal{O}\left(\delta_\beta^2\right) \nonumber\\
&& {}=\pi_r^A-\frac{1}{2} \pi_a^A-i K_\beta^A\left[\pi_r\right]+\mathcal{O}\left(\pi_a \partial, \partial^2, \pi_a^2\right).
\end{eqnarray}
The second equality follows from further expansion of exponential map. It can be seen that the higher-order contributions here have two origins: one is the higher-power operations of the thermal translation operator $\delta_{\beta}$, and the other is the higher-order terms of the $a$-type fields contained in the expansion of the exponential. One thing should also be emphasized that higher-power operations, say $\delta^n_{\beta}$, rely on specific choice of connection defined on coset manifold. 

To simplify, we keep only leading contribution and the logarithmic map gives us
\begin{equation}
    \log _{\pi_r} \tilde{\pi}_2\approx-\frac{1}{2} \pi_a-i K_\beta\left[\pi_r\right]+\mathcal{O}\left(\pi_a \partial, \partial^2, \pi_a^2\right).
\end{equation}
This is just equivalent to classical limit commonly used in the literature \cite{Glorioso:2017fpd,Haehl:2016pec,Haehl:2016uah}. Consequently, our desired KMS-shifted field reads
\begin{equation}
    \pi_{a, \beta}\approx\pi_a+i K_\beta\left[\pi_r\right],
\end{equation}
which gives us the inhomogeneous transformation rule of $a$-type fields. 

At present, we only discuss the thermal translation part and full DKMS also includes anti-unitary operation part. Remember that $a$-type field $\pi_a$ is interpreted as tangent vector, the push-forward map $\tau_{\ast}$ is needed. Then the full DKMS transformation is written as
\begin{equation}
\widetilde{\pi}_r(\vartheta x)=\tau\left(\pi_r(x)\right),
\end{equation}
\begin{equation}
    \widetilde{\pi}_a(\vartheta x)=\tau_{*, \pi_r(x)}\left[\pi_a(x)+i K_\beta\left[\pi_r\right](x)\right],
\end{equation}
with higher-order contributions neglected. Locally, this can be also written as corresponding component expression
\begin{eqnarray}
&& \widetilde{\pi}_a^A(\vartheta x)=\left(\tau_*\right)^A{ }_B\left(\pi_r(x)\right)\Bigl[\pi_a^B(x) \nonumber\\
&& {}+i\beta^\mu D_\mu \pi_r^B(x)+i \delta_{\Lambda_\beta} \pi_r^B(x)\Bigr].
\end{eqnarray}
The commonly used form of DKMS can be obtained by choosing $\beta^{\mu}=(1,0,0,0)$ and $\Lambda_{\beta}=0$ which physically corresponds to zero density system and local rest frame. One can easily check that the DKMS transformation derived here returns to the identity when applied twice. 

We close this section with several comments. In essence, DKMS is a non-local transformation. The two fields connected by the DKMS transformation are separated by a finite interval on the thermal circle. Hence, the complete quantum DKMS transformation necessitates contributions at infinite derivative order. If only the DKMS condition in the classical limit is employed in the construction of the effective field theory, the validity of such theoretical framework will be severely restricted. Although the use of the quantum DKMS condition in constructing effective theories is essential and evident, its complexity has so far hindered a thorough treatment in the literature. Recently, \cite{Firat:2025upx} has tried to implement the DKMS symmetry at all orders in $\hbar \omega/T$. 

\subsection{Transgression construction of Berry term}

As previously reviewed, we know the most important structure of type-B Goldstone is the existence of Berry term, or equivalently, symplectic structure. In this section, we attempt to generalize such interesting structure to the SK contour. 

A naive way to do such thing is to double Berry potential directly. More precisely, we may write
\begin{eqnarray}
&& S_B^{\mathrm{S} K}=S_B\left[\pi_1\right]-S_B\left[\pi_2\right] \nonumber\\
&& {}=\int \mathrm{d} t\left[A_A\left(\pi_1\right) \dot{\pi}_1^A-A_A\left(\pi_2\right) \dot{\pi}_2^A\right].
\label{eq:naive generalization}
\end{eqnarray}
After transforming into Keldysh basis, Eq. (\ref{eq:naive generalization}) can be further written as
\begin{equation}
    \label{eq:leading naive term}
    S_B^{\mathrm{S} K}=\int \mathrm{d} t\  \pi_a^A \Omega_{A B}\left(\pi_r\right) \dot{\pi}_r^B+O\left(\pi_a^3\right),
\end{equation}
with $\Omega_{A B}\left(\pi_r\right)=\partial_A A_B-\partial_B A_A$ the corresponding Berry curvature. Although this form appears to make sense and will be treated as a benchmark of subsequent construction, this approach still suffers from a number of drawbacks. 

Mathematically, the Berry potential is not globally defined on the manifold. It depends on the specific choice of coordinate patch. Moreover, this potential also shows gauge-dependence. To avoid these drawbacks, a more appropriate geometric data to construct Berry term in effective action is the Berry curvature $\Omega$. This quantity is globally well-defined on the whole manifold and gauge-independent. Therefore, we use Berry curvature as fundamental object to construct corresponding effective action terms in this section.

The geometric data at hand is then the Berry curvature. Note that we need a 1-form field in order to construct corresponding effective action. However, Berry curvature is an obvious 2-form field. Thus, we have to find a way to bridge this gap. Fortunately, a mathematical method which is called transgression has already been developed \cite{Bott:1982,Nakahara:2003}. Accordingly, we will use this method to construct Berry term in effective action.

To simplify the form of the equations that follow, we rewrite the variables in the near-diagonal region using the following notation
\begin{equation}
r^A \equiv \pi_r^A, \quad \xi^A \equiv \pi_a^A.
\end{equation}
It should be emphasized here that we have not applied any mathematical manipulation to the original variables. The sole purpose is to dispense with the $r$, $a$ subscript in the equations that follow. Next, we choose a torsion-free auxiliary connection $\nabla$ on tangent bundle $T\mathcal{M}$, with corresponding Christoffel symbols denoted by ${\Gamma^A}_{BC}={\Gamma^A}_{CB}$. This connection is not an additional physical datum. It is used to define exponential map in the near-diagonal region. Although different choices of auxiliary connection can affect the local expression of Berry terms, the exact transgression functional which will be introduced shortly is independent of the choice of connection. As a result, we believe that any physical quantities should not rely on a particular connection. This is precisely why we use the term "auxiliary" here. 

To resolve the issue that the degree of the form field exceeds the dimension of the integration, we will introduce an additional auxiliary parameter $s$, thereby increasing the dimension of the integral by one. The basic idea of this procedure is very similar to that behind the well-known Wess-Zumino-Witten (WZW) term \cite{Witten:1983tw,WESS197195}. We define the following interpolation field
\begin{equation}
\Pi(s)=\exp _r(s \xi), \quad s \in[-1 / 2,1 / 2] .
\end{equation}
This field just define a strip region. Further, compared with Eq. (\ref{eq:exponential parameterization}), the two SK contour fields are obtained as the endpoints of interpolation field
\begin{equation}
\pi_1=\Pi(1 / 2), \quad \pi_2=\Pi(-1 / 2).
\end{equation}
By using the near-diagonal geometry, we know that the following initial conditions hold
\begin{equation}
\Pi^A(0)=r^A,\left.\quad \frac{\mathrm{~d} \Pi^A}{\mathrm{~d} s}\right|_{s=0}=\xi^A .
\end{equation}
Also, we require $\partial_s\Pi$ to be parallel transported along such strip, i.e.,
\begin{equation}
    \label{eq:geodesic equation}
    \nabla (\partial_s\Pi)=0\quad \Longleftrightarrow \quad \frac{\mathrm{d}^2 \Pi^A}{\mathrm{~d} s^2}+\Gamma_{B C}^A \frac{\mathrm{d} \Pi^B}{\mathrm{~d} s} \frac{\mathrm{~d} \Pi^C}{\mathrm{~d} s}=0,
\end{equation}
which is exactly a geodesic equation. This condition is not mandatory. It is merely a convenient choice made to facilitate the subsequent derivation.   

Based on the necessary preparations made above, the following exact transgression can be defined as
\begin{equation}
    \label{eq:exact transgression}
    I_B^{S K}=\int \mathrm{d} t\  \mathrm{d}^d x \int_{-1 / 2}^{1 / 2} \mathrm{d} s\  \Omega_{A B}(\Pi) \partial_s \Pi^A \partial_t \Pi^B.
\end{equation}
The only geometric data involved in this construction is Berry curvature, which is gauge independent and irrelevant to the choice of local coordinates. Thus, Eq. (\ref{eq:exact transgression}) is globally well-defined on the whole manifold. Accordingly, we think this exact transgression is the desired Berry term generalized to the SK contour. One comment should be made here. There is no need to use any covariant derivative in this exact transgression. The derivatives $\partial_s\Pi$ and $\partial_t\Pi$ are already tangent vectors at $\Pi(s)$. Naturally, covariant derivatives enter only when we try to express the entire integrand as tensors at the base point $r=\pi_r$.  

To gain some experience, we can expand Eq. (\ref{eq:exact transgression}) covariantly to get some local terms of effective action. Without directly expanding exponential map around $s=0$, the following covariant derivation use parallel transport along the geodesic.

Let $P_s: T_r\mathcal{M}\rightarrow T_{\Pi(s)}\mathcal{M}$ denote parallel transport along the geodesic given by $\Pi(s)$, and $P^{-1}_s$ its inverse. In order to compare all tensors at a common point, we need to parallel transport them back to $T_r\mathcal{M}$. After parallel transporting Berry curvature back to $r=\Pi(s=0)$ point, one has 
\begin{equation}
P_s^{-1} \Omega_{\Pi(s)}=\Omega_r+s\left(\nabla_{\xi} \Omega\right)_r+\frac{s^2}{2}\left(\nabla_{\xi} \nabla_{\xi} \Omega\right)_r+\mathcal{O}\left(\xi^3\right).
\end{equation}
Here, the directional derivatives are defined as
\begin{equation}
\nabla_{\xi} \Omega \equiv \xi^C \nabla_C \Omega, \quad \nabla_{\xi} \nabla_{\xi} \Omega \equiv \xi^C \xi^D \nabla_C \nabla_D \Omega,
\end{equation}
which can be generalized to higher order derivatives. 

The behavior of $\partial_s\Pi$ can be directly obtained by geodesic equation. Since $\partial_s\Pi\big|_{s=0}=\xi$, we have exactly, after parallel transporting back to $r$,
\begin{equation}
P_s^{-1} \partial_s \Pi=\xi .
\end{equation}

Now we deal with the expansion of $\partial_t\Pi$. In fact, the vector field
\begin{equation}
    J(s)\equiv\partial_t \Pi
\end{equation}
is just a variation field of the geodesic family parametrized by $t$, which is usually called Jacobi field in the context of differential geometry \cite{Frankel:2011}. Thus it satisfies the corresponding Jacobi equation
\begin{equation}
\nabla_s^2 J+R\left(J, \partial_s \Pi\right) \partial_s \Pi=0,
\end{equation}
up to a sign convention of curvature. In this work, we choose to use the following convention
\begin{eqnarray}
&& \left[\nabla_C, \nabla_D\right] V^A={R^A}_{B C D} V^B, \nonumber\\
&& \quad[R(X, Y) Z]^A={R^A}_{B C D} Z^B X^C Y^D.
\end{eqnarray}
To solve this equation, we impose the initial data
\begin{equation}
J(0)=\dot{r},\left.\quad \nabla_s J\right|_{s=0}=D_t \xi,
\end{equation}
where we use $D_t$ to emphasize differentiation along the particular trajectory $r(t)$, which is defined as
\begin{equation}
D_t \xi^A=\dot{\xi}^A+\Gamma^A{ }_{B C}(r) \dot{r}^B \xi^C .
\end{equation}
Solving the Jacobi equation perturbatively gives 
\begin{equation}
P_s^{-1} \partial_t \Pi=\dot{r}+s D_t \xi-\frac{s^2}{2} R(\dot{r},\xi) \xi+\mathcal{O}\left(\xi^3\right).
\end{equation}

Inserting all these expressions into exact transgression Eq. (\ref{eq:exact transgression}), we now have local terms at the base point $r$. We can organize these terms in powers of $\xi=\pi_a$. Further, we observe that even power terms in $\xi$ are accompanied by odd powers of $s$, so they give no contribution once the $s$ integration is carried out. In other word, the Berry transgression contains only odd powers of the $a$-field. This is consistent with Eq. (\ref{eq:imaginary term}) because the Berry term is real in effective action. 

To be explicit, after restoring SK notation, we write down the concrete effective Lagrangian originating from Berry transgression up to cubic terms
\begin{equation}
    \mathcal{L}^{SK}_B=\mathcal{L}^{(1)}_B+\mathcal{L}^{(3)}_B,
\end{equation}
with
\begin{equation}
\label{eq:1st oder Berry}
\mathcal{L}^{(1)}_B=\Omega_{A B}\left(\pi_r\right) \pi_a^A \dot{\pi}_r^B 
\end{equation}
and
\begin{eqnarray}
&& \mathcal{L}^{(3)}_B=\frac{1}{12}\left(\nabla_C \Omega_{A B}\right)\left(\pi_r\right) \pi_a^C \pi_a^A D_t \pi_a^B \nonumber\\
&& {}+\frac{1}{24}\left(\nabla_C \nabla_D \Omega_{A B}\right)\left(\pi_r\right) \pi_a^C \pi_a^D \pi_a^A \dot{\pi}_r^B \nonumber\\
&& {}-\frac{1}{24} \Omega_{A B}\left(\pi_r\right) \pi_a^A {R^B}_{C D E}\left(\pi_r\right) \pi_a^C \dot{\pi}_r^D \pi_a^E.
\end{eqnarray}
It can be seen that our exact transgression construction naturally incorporates higher-order interaction terms in a self-consistent manner, without the need to introduce additional Wilsonian coefficients. Also, we see that the leading contribution is consistent with the one obtained by naive method, Eq. (\ref{eq:leading naive term}) and the higher order nonlinear terms are directly covariant. Besides, the geometrically covariant approach adopted here will make the subsequent discussion of DKMS invariance considerably simpler and more transparent.

After writing down the desired effective action, we now discuss corresponding behaviors under DKMS transformation. For thermal initial density matrix, the consistent terms in effective action should be invariant under DKMS. Hence, should the Berry term not preserve DKMS invariance, it would be ruled out in the effective field theory, which is obviously not what we seek. Fortunately, based on some reasonable assumptions, we can show that our Berry term is DKMS invariant, which is desired consequence. 

To proceed, we focus our discussion on the leading-order action $\mathcal{L}^{(1)}_B$ in this section and postpone a detailed treatment of the exact Berry transgression to Appendix \ref{app:B}. Based on the discussion of DKMS in the preceding sections, we know that the full DKMS transformation comprises a homogeneous part, which originates solely from the anti-unitary discrete operation, and an inhomogeneous part induced by the shift vector $K_{\beta}$. Firstly, we consider homogeneous part of DKMS transformation. 2-form field $\Omega$ should have definite eigenvalue under anti-unitary operation. Besides, since the anti-unitary transformation itself is a $Z_2$ symmetry, the corresponding eigenvalue can only be $\pm1$. Thus, we have
\begin{equation}
    \tau^* \Omega=\eta_{\Omega} \Omega, \quad \eta_{\Omega}= \pm 1.
\end{equation}
In some given local patch with chosen coordinates, this form can be rewritten in component:
\begin{equation}
    \Omega_{A B}(\tau(\pi))\left(\tau_*\right)^A{ }_C(\pi)\left(\tau_*\right)^B{ }_D(\pi)=\eta_{\Omega} \Omega_{C D}(\pi) .
\end{equation}
Remembering that an extra time derivative is also involved in Eq. (\ref{eq:1st oder Berry}), we denote the eigenvalue associated with time derivative as $\epsilon_{\mathrm{B}}$. Then the invariance under DKMS demands that
\begin{equation}
\label{eq:homo constraint}
    \epsilon_{\mathrm{B}} \eta_{\Omega}=1.
\end{equation}
Since an anti-unitary transformation necessarily involves the operation of time reversal, $\epsilon_{\mathrm{B}}=-1$ holds. Consequently, the eigenvalue of 2-form $\Omega$ should be $-1$, i.e., $\eta_{\Omega}=-1$.

Next, we can further discuss the inhomogeneous part of DKMS. For leading-order term, we have
\begin{eqnarray}
&& \tilde{\mathcal{L}}_{B}^{(1)}  =\epsilon_{\mathrm{B}} \Omega_{A B}\left(\tau\left(\pi_r\right)\right)\left(\tau_*\right)^A{ }_C\left(\pi_a^C+i K^C\right)\left(\tau_*\right)^B{ }_D D_t \pi_r^D \nonumber\\
&& {}=\epsilon_{\mathrm{B}}\left(\tau^* \Omega\right)_{C D}\left(\pi_a^C+i K^C\right) D_t \pi_r^D \nonumber\\
&& {}=\mathcal{L}_{B}^{(1)}+i \Omega\left(K, D_t \pi_r\right),
\end{eqnarray}
with Eq. (\ref{eq:homo constraint}) applied. Because 2-form field $\Omega$ is completely antisymmetric, there is no reason to expect a cancellation between the extra term produced here and any extra terms that may arise from totally symmetric second-rank tensors, say, the metric $g$. Accordingly, this extra term $\Omega\left(K, D_t \pi_r\right)$ should either vanish or be rewritten as total derivative. 

To proceed, we work in local rest frame and $\beta^{\mu}=\beta u^{\mu}=\beta(1,0,0,0)$. In such local frame, the shift vector $K_{\beta}$ can be parameterized as
\begin{equation}
    \label{eq:local shift vector}
    K_\beta=\beta D_t \pi_r+\nu^I k_I\left(\pi_r\right),
\end{equation}
with $k_I(\pi_r)$ the tangent vector on coset manifold, generated by internal symmetry. Similar to $\beta$, $\nu^I$ introduced above can be treated as external parameters corresponding to thermal twist, which are analogous to chemical potentials.  

It can be readily seen that the contribution from the first term of the Eq. (\ref{eq:local shift vector}) vanishes
\begin{equation}
    \Omega\left(\beta D_t \pi_r, D_t \pi_r\right)=0.
\end{equation}
The thermal twist part, however, requires a more detailed discussion. For the sake of completeness, before analyzing its contribution to DKMS invariance, we first examine the correct behavior of this part under the DKMS transformation. 

According to its nature as a tangent vector, the behavior of $k_I$ under the anti-unitary operation is given by
\begin{equation}
    \label{eq:thermal twist push-forward}
    \tau_* k_I(\pi_r)=\rho_I{ }^J k_J(\tau(\pi_r)).
\end{equation}
The covariance of shift vector $K_{\beta}$ further demands that
\begin{eqnarray}
&& \tau_{*, \pi(x)}\left[\nu^I(x) k_I(\pi(x))\right] \nonumber\\
&& {}=(\Theta \nu)^J(\vartheta x) k_J(\tau(\pi(x))) \nonumber\\
&& {}=\nu^I(x)\rho_I{ }^J k_J(\tau(\pi_r)) ,
\end{eqnarray}
where the last equality follows directly from applying Eq. (\ref{eq:thermal twist push-forward}). As a result, the correct DKMS transformation law for the thermal twist reads as follows:
\begin{equation}
    (\Theta \nu)^J(y)=\nu^I\left(\vartheta^{-1} y\right) \rho_I{ }^J \ \ \text{with} \ \ y=\vartheta x.
\end{equation}
If this transformation rule no longer holds, the DKMS symmetry would fail already at the level of the thermal twist sector. 

In our local rest frame, the interior product can be given by
\begin{equation}
    \label{eq:Poincare lemma}
    \iota_{k_I} \Omega=\mathrm{d} \mu_I,
\end{equation}
with $\mu_I$ a scalar function, i.e., $\mu_{I}: \mathcal{M}\rightarrow\mathbb{R}$. It should be emphasized here that Eq. (\ref{eq:Poincare lemma}) only holds in a local patch rather than entire coset manifold. The obstruction to this equation being valid globally is the de Rham cohomology $H^1(\mathcal{M})$ of the coset manifold. For a generic coset manifold, one certainly cannot simply assume that its de Rham cohomology structure is trivial. From another perspective, Eq. (\ref{eq:Poincare lemma}) is equivalent to a special choice of internal symmetry generating vector $k_I$, that is to say, the vector $k_I$ turns into a symplectic vector:
\begin{equation}
    \mathcal{L}_{k_I} \Omega=\mathrm{d} \iota_{k_I} \Omega+\iota_{k_I} \mathrm{d} \Omega=0.
\end{equation}

Based on the discussion made before, we now have
\begin{equation}
    \Omega\left(k_I, D_t \pi_r\right)=\left(d \mu_I\right)\left(D_t \pi_r\right)=D_t \mu_I(\pi_r) .
\end{equation}
Accordingly, the following equation holds:
\begin{equation}
    \nu^I \Omega\left(k_I, D_t \pi_r\right)=\nu^I D_t \mu_I\left(\pi_r\right)=D_t\left(\nu^I \mu_I\right)-\left(D_t \nu^I\right) \mu_I.
\end{equation}
If the thermal twist is assumed to be stationary, i.e., $D_t\nu^I=0$, the extra term given by DKMS transformation can be written as a total derivative. Such total derivative terms give a vanishing contribution in the sense of integration. Thus, the leading-order term $\mathcal{L}^{(1)}_B$ is DKMS invariant.  

The stationary assumption is physically reasonable. Intuitively, the behavior of non-stationary thermal twist is equivalent to some external driving source. For a driven system, one should properly incorporate external fields into effective theory. But in this work, we investigate effective theory without introducing any external sources. As a result, we can only focus on stationary thermal twist and postpone more general case to future work. 

Having completed the discussion of the leading-order, one might be tempted to proceed directly to the higher-order terms in the local expansion of transgression integral. However, the higher-order expansion involves an increasing number of terms, and the extra terms brought about by the inhomogeneous shift of the DKMS transformation are therefore highly complicated. This makes it exceedingly difficult to demonstrate that, at each order, these extra terms can combine into a total derivative. Fortunately, expansion of Berry transgression does not introduce additional Wilsonian coefficients. Thus, the higher-order terms and the leading-order term are governed by the same coefficient. We may, in fact, refrain from expanding the exact transgression integral and instead prove its DKMS invariance directly. We will give a detailed derivation in Appendix \ref{app:B}. The results demonstrate that, under certain reasonable assumptions, the Berry term constructed by transgression method in the effective theory does satisfy DKMS invariance. 

Before ending this section, here are some comments. Dynamically, the Berry term generally gives rise to precession-like behavior. Therefore, such a term does not contribute to entropy production. Furthermore, the fact that the Berry term itself satisfies DKMS invariance implies that it does not mix directly with other sectors of the effective theory. In the SK effective theory, these other sectors encompass the dissipative and noise parts. As a result, Berry term bears no direct relation to dissipation and noise. Of course, when computing correlation functions, the Berry term does exert an indirect influence on the physical observables of the dissipative system.

\subsection{The symmetric sector: noise and dissipation}

The standard coset based SK constructions organize local operators using Maurer-Cartan data and invariant tensors in a chosen coset chart or frame \cite{Akyuz:2023nbo,Landry:2019iel,Hongo:2024brb}. In the preceding discussion, we have already pointed out that the $a$-type fields are essentially tangent vectors based at moving point $\pi_r(x)$. However, a formulation based only on local components must therefore keep track of changes of chart, frame and base point on coset manifold when different vectors are compared. This motivates us to organize the SK effective theory directly in terms of global $G$-invariant tensor fields on $\mathcal{M}$ and pulling them back by $\pi_r$, where $G$ is the full symmetry group of our system. Denote the unbroken subgroup as $H$. Then, on a homogeneous space, $G$-invariant tensors and $H$-invariant tensors at a given reference point contain equivalent algebraic information, so this reorganization introduces no additional Wilsonian couplings.

For type-B system, the universal geometric data is the Berry curvature $\Omega$, which has already been investigated in previous discussion. For the symmetric sector, we also have a positive $G$-invariant metric tensor $g$. Accordingly, two tensors naturally generate
\begin{equation}
J^A{ }_B \equiv g^{A C} \Omega_{C B}, \quad \mathcal{G}_{A B} \equiv \Omega_{C A} g^{C D} \Omega_{D B}=-\left(g J^2\right)_{A B} .
\end{equation}
The new tensor $\mathcal{G}$ is symmetric, globally $G$-invariant and positive semi-definite. Also, it is nondegenerate when $\Omega$ is nondegenerate. It should be emphasized that this new tensor $\Omega$ is not always independent of metric $g$. If $J^2=-\lambda^2\mathbf{1}$, this means that $J$ is an almost complex structure after a suitable normalization, then $\mathcal{G}=\lambda^2 g$. Otherwise, $g$ and $\mathcal{G}$ are linearly independent. Thus, they admit independent Wilsonian coefficients and provide independent $G$-invariant tensors for effective operator basis. Under anti-unitary operation, we have $\tau^{\ast}\mathcal{G}=\mathcal{G}$ because of $\tau^{\ast}\Omega=-\Omega$ and $\tau^{\ast}g=g$.

Before proceeding further with the construction of effective action, we need to first discuss certain geometric structures. To be more precise, the $a$-type field defined on physical spacetime is obtained by pulling back tangent vector $\pi_a\in T_{\pi_r(x)}\mathcal{M}$ along the map $\pi_r:X\rightarrow\mathcal{M}$. Thus, the physical configuration defines the following pullback bundles
\begin{equation}
E_r \equiv \pi_r^* T \mathcal{M}, \quad E_r^* \equiv \pi_r^* T^* \mathcal{M} .
\end{equation}
For notational compactness, we also set $P^{(1)}=g$ and $P^{(2)}=\mathcal{G}$. Thus, the Berry curvature and these symmetric invariant tensors just define following maps
\begin{equation}
\Omega, P^{(\ell)}: E_r \longrightarrow E_r^*.
\end{equation}
Further, composing these maps with pullback derivatives produces local differential operators
\begin{equation}
\mathbb{O}\left[\pi_r\right]: \Gamma\left(E_r\right) \longrightarrow \Gamma\left(E_r^*\right) .
\end{equation}
Here $\Gamma(E_r)$ denotes the space of smooth sections of the pullback bundle $E_r$. The mathematical meaning of $\Gamma(E^{\ast}_r)$ is analogous, except that the bundle is replaced by $E^{\ast}_r$. Also, this local differential operator may depend on $\pi_r$, on the derivatives of $\pi_r$ retained at the chosen EFT order, and on external background sources and their derivatives. Accordingly, if acting this local differential operator on a vector $v\in \Gamma(E_r)$, we have the following expression in component form
\begin{eqnarray}
&& (\mathbb{O} v)_A=\sum_{m=0}^N \mathcal{O}_{A B}^{\mu_1 \cdots \mu_m}\left(\pi_r, D \pi_r, \ldots ; \text { sources }\right) \nonumber\\
&& D_{\mu_1} \cdots D_{\mu_m} v^B,
\end{eqnarray}
with suitable covariant derivative defined as
\begin{equation}
\label{eq:covariant derivative}
D_\mu v^A=\partial_\mu v^A+\Gamma^A{ }_{B C}\left(\pi_r\right) D_\mu \pi_r^B v^C.
\end{equation}
For the convenience of subsequent presentation, we further define: $U^A_i\equiv D_i\pi^A_r$. 

Moreover, we note that near-diagonal geometry supplies precisely the relevant sections,
\begin{equation}
\label{eq:covectors}
\pi_a, \  D_t \pi_r, \  K_\beta\left[\pi_r\right] \in \Gamma\left(E_r\right).
\end{equation}
Besides, we also observe that the functional derivative of free energy density $f$ with respect to $\pi_r$, which can be denoted as $\mathcal{E}_A(f)[\pi_r]$, is also a element in $\Gamma(E^{\ast}_{r})$. It is not a linear operator acting on one of the vectors in Eq. (\ref{eq:covectors}). Therefore, this part contributes to an independent physical sector, which will be referred as conservative sector hereinafter. 

To get some intuition, the three different deterministic responses which contributes at linear order in the expansion of the $a$-type fields are
\begin{eqnarray}
&& \mathcal{R}_{\mathrm{B}, A}\left[\pi_r\right] \equiv \Omega_{A B} D_t \pi_r^B, \nonumber\\
&& \ \mathcal{R}_{\mathcal{F}, A}\left[\pi_r\right] \equiv-\mathcal{E}_A(f)\left[\pi_r\right], \nonumber\\
&& \ \mathcal{R}_{\mathrm{diss}, A}\left[\pi_r\right] \equiv-T\left(\mathbb{L} K_\beta\right)_A\left[\pi_r\right].
\end{eqnarray}
The physical interpretation of first two parts are quite clear and the third one needs more clarification. Based on the structure of SK effective theory review before, we know the third part is explained as dissipative response. We introduce an extra differential operator $\mathbb{L}$ here and this will be proved useful when discussing DKMS invariance of such part. Besides, it should be emphasized that this differential operator $\mathbb{L}$ is understood to be functional of $\pi_r$.  Although these three parts have entirely distinct physical interpretations, they admit a unified description in the language of bundles at the level of mathematical construction. To simplify the corresponding expression, we can introduce the following notation 
\begin{eqnarray}
&& \langle u, \alpha\rangle \equiv \int_X \mathrm{~d}^{d+1} x\ u^A \alpha_A \nonumber\\
&& \ \Longrightarrow \  S_{\mathrm{det}}^{(1)}=\left\langle\pi_a, \mathcal{R}_{\mathrm{B}}+\mathcal{R}_{\mathcal{F}}+\mathcal{R}_{\mathrm{diss}}\right\rangle,
\end{eqnarray}
with $u\in \Gamma (E_r)$ and $\alpha\in \Gamma (E^{\ast}_r)$. 

Before proceeding to discuss specific operator classification, one subtlety should be clarified. As was indicated previously, $\mathcal{E}_A(f)[\pi_r]$ corresponding to conservative sector is a nonlinear map, whereas $\mathbb{L}[\pi_r]$ describing the dissipative part is a linear differential operator. Thus, to compare the conservative sector with dissipative sector properly, we need to linearize the nonlinear map around some chosen background configuration. For $\pi_r(\epsilon)=\exp_{\pi_r}(\epsilon v)$ where $v\in \Gamma(E_r)$, define
\begin{equation}
\left.\left(\mathbb{H}_f v\right)_A \equiv \frac{D}{D \epsilon}\right|_{\epsilon=0} \mathcal{E}_A(f)\left[\pi_r(\epsilon)\right].
\end{equation}
Taking into account the generally curved coset manifold, a covariant derivative along a curve $\pi_r(\epsilon)$ is employed in this definition. The physical meaning of this new operator $\mathbb{H}_f$ is quite simple. It governs small fluctuations of the conservative force about the chosen $r$-type field configuration. In other word, this is simply the functional Hessian of the free energy density. Moreover, symmetry of the second variation implies $\mathbb{H}^{\dagger}_f=\mathbb{H}_f$. On the contrary, as will be seen in the subsequent discussion, $\mathbb{L}$ is constrained by both positivity and DKMS invariance. 

For the simplest case, namely Gaussian fluctuations, we can write down the following quadratic form $i T\left\langle\pi_a, \mathbb{L} \pi_a\right\rangle$. Similarly, higher fluctuation contributions can be described by local covariant $n$-point kernels. Accordingly, the derivative expansion and the expansion in the $a$-type field usually used in organizing SK effective field theory amount to classifying local operators with the required coset manifold indices. Further, geometry determines the admissible index contractions.  

After finishing the discussion about general geometric structures, we now consider concrete operator classification at fixed order. Based on the general structures discussed before, a local term can be constructed by contracting global invariant coset tensors evaluated on $\pi_r$, spatial tensors such as $\delta^{ij}$, suitable covariant derivatives, and the possible allowed external fields. Besides, as far as effective field theory is concerned, there is one point that deserves special emphasis, namely: operators related by integration by parts, possible coset space identities or perturbative field redefinitions are redundant. This point gives rise to some technical complications in the practical construction of generic effective field theories and needs careful treatment. Typically, we can write a general local $n$-point kernel in the following compact way 
\begin{eqnarray}
&& \mathcal{C}_n\left(u_1, \ldots, u_n\right)=  \sum_{\substack{m_1, \ldots, m_n, q \geq 0 \\
|\boldsymbol{m}|+q \leq N}} \int_X \mathrm{~d}^{d+1} x\ \nonumber\\
&& C_{A_1 \cdots A_n}^{\boldsymbol{\mu}_1 \cdots \boldsymbol{\mu}_n ; q}\left(\pi_r ; U, D U, \ldots\right)  \prod_{s=1}^n\left(D_{\boldsymbol{\mu}_s} u_s\right)^{A_s} .
\label{eq:kernel basis}
\end{eqnarray}
Here $u_s\in \Gamma(E_r)$, with $s=1,\dots,n$. In our SK effective action, these can be $\pi_a$, $K_{\beta}$, or linear combinations of them. Also, to avoid potential obstacles in understanding, we emphasized that the indices $A_s=1,\dots, {\rm{dim}} \mathcal{M}$ are coset manifold indices at $\pi_r(x)$. 

Due to its rather compact form, Eq. (\ref{eq:kernel basis}) requires further clarification. For each field, $\boldsymbol{\mu}_s$ denotes the ordered list of spacetime indices carried by the derivatives, i.e.,
\begin{eqnarray}
&& \boldsymbol{\mu}_s=(\mu_{s,1},\dots,\mu_{s,m_s}), \nonumber\\
&& \quad \text{and} \quad
  D_{\boldsymbol{\mu}_s}u_s
  \equiv D_{\mu_{s,1}}\cdots D_{\mu_{s,m_s}}u_s .
\end{eqnarray}
Thus $m_s\geq0$ is the number of covariant derivatives acting directly on $u_s$. The collective notation
$\boldsymbol{m}=(m_1,\ldots,m_n)$ denotes these derivative numbers, and $|\boldsymbol{m}|\equiv\sum_{s=1}^n m_s$ just gives us the total number of derivatives acting on the inserted fields. Here $D_\mu$ is the coset manifold covariant
derivative previously defined in Eq. (\ref{eq:covariant derivative}). Here, we demand  $\boldsymbol{\mu}_s$ to be ordered because covariant derivatives need not commute on a general $r$-type field configuration.  

We interpret the remaining unspecified parameter $q$ in Eq. (\ref{eq:kernel basis}) as the derivative order carried by the coefficient tensor $C_{A_1 \cdots A_n}^{\boldsymbol{\mu}_1 \cdots \boldsymbol{\mu}_n ; q}$ itself. Thus, $q$ should be a non-negative integer. Note that such coefficient tensor relies on $U_i^A \equiv D_i \pi_r^A$ defined previously, higher derivatives acting on $U$, and also possible external sources and their derivatives. Accordingly, in our present counting, each factor of $U$ carries one derivative, and every additional derivative $D$ acting on $U$ just increases the derivative order by one. Consequently, the constraint $|\boldsymbol{m}|+q\leq N$ just enforces the maximum total derivative order to be $N$. Further, we need to be emphasize one more thing. The coefficient tensor $C_{A_1 \cdots A_n}^{\boldsymbol{\mu}_1 \cdots \boldsymbol{\mu}_n ; q}$ must be invariant under the global $G$ action. As can be seen, Eq. (\ref{eq:kernel basis}) gives a finite operator basis before any concrete relations about coset manifold are imposed. If we further specify coset-dependent curvature relations and identities which are suitable for the investigation of specific models, this general operator basis reduces to the basis used in actual model building. 

To simplify, we now specialize the general kernel Eq. (\ref{eq:kernel basis}) to the symmetric quadratic sector relevant to Gaussian fluctuations and, below, to the symmetric dissipative response. Thus, we need to set $n=2$, retain only spatial derivatives, and also truncate at $N=4$. Namely, we have the local operator $\mathbb{O}$, satisfying  
\begin{equation}
\label{eq:quadratic ansatz}
\mathcal{C}_2(u, v)=\langle u, \mathbb{O} v\rangle = \int_X \mathrm{~d}^{d+1} x\ u^A(\mathbb{O} v)_A.
\end{equation}
Note Eq. (\ref{eq:quadratic ansatz}) is Eq. (\ref{eq:kernel basis}) at $(n,N)=(2,4)$, not a separate ansatz. To enumerate the terms with $|\boldsymbol{m}|+q\leq4$, we assume the parity of our theory is even. Thus, the possible effective field theory terms contain only an even number of spatial derivatives. Let the tensors $V$, $M$ and $Q$ collect, respectively, the terms with zero, two and four derivatives. It follows that the local differential operator $\mathbb{O}$ can be written as
\begin{eqnarray}
&& (\mathbb{O} v)_A=V_{A B} v^B-D_i\left(M_{A B}^{i j} D_j v^B\right) \nonumber\\
&& {}+D_i D_j\left(Q_{A B}^{i j ; k l} D_k D_l v^B\right),
\label{eq:general operator}
\end{eqnarray}
with
\begin{equation}
V_{A B}=V_{B A}, \quad M_{A B}^{i j}=M_{B A}^{j i}, \quad Q_{A B}^{i j ; k l}=Q_{A B}^{(i j) ;(k l)}=Q_{B A}^{k l ; i j} .
\end{equation}
Here, $(ij)$ refers to standard symmetrization of two indices $i$ and $j$. These relations precisely originate from our definition of kernel $\mathbb{O}$. Based on our present truncation, i.e., $|\boldsymbol{m}|+q\leq 4$, these operators can be organized as
\begin{eqnarray}
&& V=V^{(0)}+V^{(2)}+V^{(4)}, \nonumber\\
&& \quad M=M^{(0)}+M^{(2)}, \quad Q=Q^{(0)}.
\end{eqnarray}
For example, $M^{(2)}$ contains two derivatives in its coefficient, while $Q^{(0)}$ contains none in its coefficient. 

Now we construct $q=0$ coefficient tensors firstly. Remember that we have already set $P^{(1)}_{AB}=g_{AB}$ and $P^{(2)}_{AB}=\mathcal{G}_{AB}$ before. We will retain this convention in the subsequent derivation. We further assume our system is rotationally invariant. As a result, the rank-2 spatial tensor available for contraction is the $\delta^{ij}$. Thus, we have the following coefficient tensors, written as
\begin{eqnarray}
&& V_{A B}^{(0)}  =\sum_{\ell=1}^2 v_{\ell} P_{A B}^{(\ell)}, \nonumber\\
&& M_{A B}^{(0) i j}  =\sum_{\ell=1}^2 m_{\ell} P_{A B}^{(\ell)} \delta^{i j}, \nonumber\\
&& Q_{A B}^{(0) i j ; k l}  =\sum_{\ell=1}^2 P_{A B}^{(\ell)}\left(r_{\ell 1} \delta^{i j} \delta^{k l}+r_{\ell 2} \delta^{i(k} \delta^{l) j}\right) .
\label{eq:1st order operator}
\end{eqnarray}
Here, $v_{\ell}$, $m_{\ell}$ and $r_{\ell a}$ are corresponding Wilsonian coefficients. 

Similarly, we can further construct the coefficient tensors for the case $q=2$. The structure in this case is considerably more involved than that for $q=0$. They contain two derivatives of the $r$-type field. To proceed, we can define the following tensors
\begin{eqnarray}
&& X_{\ell}  \equiv P_{A B}^{(\ell)} U_i^A U^{i B},\ \nonumber\\
&& Z_{A B}^{\ell m}  \equiv P_{C(A}^{(\ell} P_{B) D}^{m)} U_i^C U^{i D}, \nonumber\\
&& \ W_{A B}^{\ell m, i j} \equiv P_{C(A}^{(\ell} P_{B) D}^{m)} U^{(i C} U^{j) D}.
\end{eqnarray}
Here we have assumed that the only globally invariant tensors are of rank two. If the coset manifold admits a globally invariant rank-4 tensor, which may be denoted by $C^{(s)}_{ABCD}$ where $s$ labels the independent tensors, then some extra tensors can also be written down, say $Y_{A B}^s \equiv C_{(A|C| B) D}^{(s)} U_i^C U^{i D}$. In this work, we shall not discuss such more complicated situations; rather, we leave them for future investigation. Next, we contract these tensors with the available rank-2 tensors and this procedure gives us
\begin{eqnarray}
&& V_{A B}^{(2)}  =\sum_{\ell, m} a_{\ell m} X_{\ell} P_{A B}^{(m)}+\sum_{\ell \leq m} b_{\ell m} Z_{A B}^{\ell m}, \nonumber\\
&& M_{A B}^{(2) i j}  =\delta^{i j}\left[\sum_{\ell, m} \bar{a}_{\ell m} X_{\ell} P_{A B}^{(m)}+\sum_{\ell \leq m} \bar{b}_{\ell m} Z_{A B}^{\ell m}\right] \nonumber\\
&& {}+\sum_{\ell \leq m} d_{\ell m} W_{A B}^{\ell m, i j} .
\label{eq:2nd order operator}
\end{eqnarray}

Further, we can construct the explicit expression of $V^{(4)}$. Its coefficients are generated by products of $X_{\ell}$, $Z^{\ell m}$, together with $D_i U^{iA}$ and $D_iU^A_j$. Although the procedure of construction is analogous, its explicit form is considerably more complicated and will therefore not be written out here. It should be emphasized that further independent invariant tensors must be added whenever they are allowed. 

We stress that only the available geometric data have been used here to construct all admissible terms. To get the final SK effective action, some more physical constraints are needed. Especially, the number of independent Wilsonian coefficients is reduced by DKMS invariance, which enforces nontrivial relations between terms originating from different physical sectors. We now turn to this point.  

Previous studies have shown that, in the SK effective field theory, the fluctuating and dissipative sectors appear in combination \cite{Crossley:2015evo,Glorioso:2020loc,Donos:2023ibv,Hongo:2024brb}. Still, we choose to work in a stationary local rest frame, with $T=\beta^{-1}$. As a result, the most general local quadratic action can be written as
\begin{equation}
S_{\text {diss }+ \text { fluc }}^{(2)}=T\left[-\left\langle\pi_a, \mathbb{D} K_\beta\left[\pi_r\right]\right\rangle+i\left\langle\pi_a, \mathbb{N} \pi_a\right\rangle\right].
\end{equation}
For later convenience, we shall call $\mathbb{D}$ the response kernel and $\mathbb{N}$ the noise kernel. Note that SK reality, namely, Eq. (\ref{eq:imaginary term}), requires $\mathbb{N}^{\dagger}=\mathbb{N}$. Also, Eq. (\ref{eq:convergence of path integral}) constrains the operator $\mathbb{N}$ to be positive semi-definite, i.e., $\left\langle v,\mathbb{N}v \right\rangle\geq 0$ in the domain of validity of the effective theory. 

The detailed properties of DKMS transformation have been discussed in the previous section. Here, we only need to further define the following DKMS transformed covariant operator
\begin{equation}
\label{eq:DKMS transformed operator}
\mathbb{O}^{\Theta}\left[\pi_r ; \nu\right] \equiv \tau^* \circ \mathbb{O}\left[\tau \circ \pi_r ; \vartheta_* D, \Theta \nu\right] \circ \tau_*,
\end{equation}
with $(\Theta \nu)^J(y)=\nu^I\left(\vartheta^{-1} y\right) \rho_I^J$ which has been shown previously. Although Eq. (\ref{eq:DKMS transformed operator}) seems quite abstract, its physical interpretation is rather simple. It means that both coset space indices are rotated by $\tau$, all derivatives are reflected by $\vartheta$, and the possible thermal twist is replaced by $\Theta \nu$. Consequently, after the change of integration variables to their $\vartheta$ transformed counterparts, we have
\begin{equation}
\left\langle\tau_* u, \mathbb{O}\left[\tau \circ \pi_r ; \vartheta_* D, \Theta \nu\right] \tau_* v\right\rangle=\left\langle u, \mathbb{O}^{\Theta} v\right\rangle .
\end{equation}

Based on above preparation, we now discuss the response kernel and noise kernel in detail. For the response kernel, microscopic time reversal interchanges the perturbation and response legs. The retarded kernel is therefore mapped to the advanced kernel. In the notation above this statement is $\left(\mathbb{D}^{\Theta}\right)_{A B}(x, y)=\mathbb{D}_{B A}(y, x)$. For a local differential operator, this is precisely the formal adjoint. Thus, we have the following relation
\begin{equation}
\left\langle u, \mathbb{D}^{\Theta} v\right\rangle=\langle v, \mathbb{D} u\rangle=\left\langle u, \mathbb{D}^{\dagger} v\right\rangle .
\end{equation}
On the contrary, the noise kernel is the coefficient of a pair of identical $a$-type fields and corresponds to the symmetrized fluctuation kernel. It should be even under the same microscopic reversal. In other word, $\left(\mathbb{N}^{\Theta}\right)_{A B}(x, y)=\mathbb{N}_{A B}(x, y)$. Combining with SK reality, we thus have 
\begin{equation}
\left\langle u, \mathbb{N}^{\Theta} v\right\rangle=\langle u, \mathbb{N} v\rangle=\langle v, \mathbb{N} u\rangle .
\end{equation}
Since these equalities hold for arbitrary $u$ and $v$, the following operator relations hold
\begin{equation}
\mathbb{D}^{\Theta}=\mathbb{D}^{\dagger}, \quad \mathbb{N}^{\Theta}=\mathbb{N} .
\end{equation}
Physically, these are the well-known Onsager constraints associated with the anti-unitary operation. In the following, we further discuss the inhomogeneous shift part of DKMS transformation.

Applying all the relations obtained before, the terms that remain after subtracting the original action are
\begin{eqnarray}
&& \widetilde{S}_{\mathrm{diss}+\text { fluc }}^{(2)}-S_{\mathrm{diss}+\text { fluc }}^{(2)}=T\Bigl[2\left\langle\pi_a,\left(\mathbb{D}_{\text {sym }}-\mathbb{N}\right) K_\beta\right\rangle \nonumber\\
&& {}+i\left\langle K_\beta,\left(\mathbb{D}_{\text {sym }}-\mathbb{N}\right) K_\beta\right\rangle  \Bigr],
\end{eqnarray}
with $\mathbb{D}_{\text {sym }}=(\mathbb{D}+\mathbb{D}^{\dagger})/2$. Since $\pi_a$ and $K_{\beta}$ are independent local arguments in this identity, the DKMS invariance requires
\begin{equation}
\label{eq:classical FDT}
    \mathbb{N}=\mathbb{D}_{\text{sym}}\equiv \mathbb{L},\quad \mathbb{L}^{\dagger}=\mathbb{L}\geq 0, \quad \mathbb{L}^{\Theta}=\mathbb{L}.
\end{equation}
Physically, Eq. (\ref{eq:classical FDT}) gives us local classical fluctuation-dissipation relation at operator level. Here, we emphasize "classical" because we have not employed the full quantum DKMS transformation. 

Within the geometric construction scheme adopted in the present work, these general operators, $\mathbb{D}_{\text{sym}}$ and $\mathbb{N}$, can be expressed in terms of the operator basis discussed and constructed earlier. That is,
\begin{equation}
    \mathbb{D}_{\text{sym}}=\sum_q d_q\mathbb{O}_q, \quad \mathbb{N}=\sum_q n_q\mathbb{O}_q.
\end{equation}
Accordingly, the fluctuation-dissipation relation just gives us a very simple constraint $n_q=d_q$. Moreover, if we assume $\mathbb{O}^{\Theta}_q=\sigma_q \mathbb{O}_q$ with $\sigma_q=\pm 1$, the associated Wilsonian coefficient, collectively denoted as $c_q(\nu)$, thus obeys
\begin{equation}
c_q(\nu)=\sigma_q c_q(\Theta \nu) .
\end{equation}
Remembering that the tensors $g$ and $\mathcal{G}$ are $\Theta$ even, so the Wilsonian coefficients of these even operators should be even under thermal twist operation. 

Finally, we discuss the remaining sector, which is the conservative response from the free energy density, and corresponding constraints originating from requirement of DKMS invariance. We can organize the static free energy density by using the spatial derivative expansion, i.e., $f=f_0+f_2+f_4+\cdots$. The leading term sensitive to a non-uniform Goldstone configuration is the two derivative term which can be written as
\begin{equation}
f_2 \equiv \frac{1}{2} \kappa_{A B}[\pi_r] U_i^A U^{i B}, \quad \kappa_{A B}=\kappa_g g_{A B}+\kappa_{\Omega} \mathcal{G}_{A B}.
\end{equation}
Further, taking functional derivative with respect to $\pi_r$ gives us desired covector 
\begin{equation}
\mathcal{E}_A\left(f_2\right) \equiv \frac{\delta \mathcal{F}_2}{\delta \pi_r^A}=\frac{1}{2} \nabla_A \kappa_{B C} U_i^B U^{i C}-D_i\left(\kappa_{A B} U^{i B}\right).
\end{equation}
According to the basic structure of SK effective action, coupling this covector to $a$-type field $\pi_a$ just gives us corresponding action $S_{\mathcal{F}}$ at given order of spatial derivative. 

To obtain terms involving higher derivatives, it is convenient to use the following tensors 
\begin{eqnarray}
&& H_{i j}^{(\ell)} \equiv P_{A B}^{(\ell)} U_i^A U_j^B\sim\mathcal{O}\left(\partial^2\right), \nonumber\\
&& \quad X_{\ell}=\delta^{i j} H_{i j}^{(\ell)}, \nonumber\\
&& \quad \mathfrak{t}^A \equiv D_i U^{i A}\sim\mathcal{O}\left(\partial^2\right) ,
\end{eqnarray}
together with $U^A_i\sim \mathcal{O}(\partial)$ itself. It follows that we can have
\begin{eqnarray}
&& f=  f_0+\frac{\kappa_g}{2} X_1+\frac{\kappa_{\Omega}}{2} X_2+c_{\ell m} X_{\ell} X_m \nonumber\\
&& {}+\widetilde{c}_{\ell m} H_{i j}^{(\ell)} H^{(m) i j}+c_{\mathfrak{t}, \ell} P_{A B}^{(\ell)} \mathfrak{t}^A \mathfrak{t}^B \nonumber\\
&& {}+c_R R_{A B C D} U_i^A U_j^B U^{i C} U^{j D}+\mathcal{O}\left(\partial^6\right), \quad \ell,m=1,2.
\end{eqnarray}
Here only the symmetric parts $c_{\ell m}=c_{m\ell}$ and $\widetilde{c}_{\ell m}=\widetilde{c}_{m\ell}$ are independent.

Of course, for the conservative sector presented above to be physically meaningful, its invariance under DKMS must be verified. The static functional should be required to be $\Theta$ even. To clarify, the following relations hold
\begin{eqnarray}
&& \mathcal{F}^{\Theta}\left[\pi_r ; \nu\right] \equiv \mathcal{F}\left[\tau \circ \pi_r ; \Theta \nu\right]  =\mathcal{F}\left[\pi_r ; \nu\right], \nonumber\\
&& \tau^* \mathcal{E}(f)\left[\tau \circ \pi_r ; \vartheta_* D, \Theta \nu\right]  =\mathcal{E}(f)\left[\pi_r ; D, \nu\right] .
\end{eqnarray}
To get second relation, the fact that $\mathcal{E}(f)$ is a covector has been used. Then, we need to check the inhomogeneous shift part. It can be easily shown that
\begin{equation}
\widetilde{S}_{\mathcal{F}}^{(1)}-S_{\mathcal{F}}^{(1)}=-i \int_X \mathrm{~d}^{d+1} x K_\beta^A \mathcal{E}_A(f)=-i \int \mathrm{~d} t \delta_\beta \mathcal{F} .
\end{equation}
As can be seen, the extra term originates from thermal variation of free energy. After discarding all possible boundary terms, we have $\delta_\beta \mathcal{F}=\beta \partial_t \mathcal{F}+\delta_{\Lambda_\beta} \mathcal{F}$. Because $\mathcal{F}$ is constructed from $G$-invariant data, its internal variation should vanish, i.e., $\delta_{\Lambda_{\beta}}\mathcal{F}=0$. Accordingly, the extra term is only a boundary term and can be discarded by suitable boundary conditions.

One may notice that our discussion and formulation are restricted to the case of Gaussian noise, i.e., only the quadratic order in the expansion of the $a$-type fields is taken into account. Although our construction scheme can yield tensors and operators suited for non-Gaussian fluctuations, the discussion of DKMS invariance for such terms is problematic. The full quantum DKMS transformation is highly nonlinear. For technical simplicity, in this work we retain only the leading-order contribution from the inhomogeneous shift of $a$-type field. However, when we attempt to discuss non-Gaussian terms, the contributions from the nonlinear part of the DKMS transformation can no longer be simply ignored. It is to be expected that this will impose stronger constraints on the nonlinear terms in the effective field theory. While non-Gaussian fluctuations have been addressed in the literature within SK formalism \cite{Lin:2023bli}, the case of an initial thermal state has not been treated, meaning that the DKMS invariance constraint is abandoned. Therefore, a satisfactory discussion of non-Gaussian fluctuations in thermal systems calls for a deeper and more systematic treatment of the full quantum DKMS symmetry. We hope this can be done in the future work.  
 
\subsection{Comparison with usual coset construction}

In this section, we briefly compare the commonly used coset construction scheme with the approach adopted in this work. 

The usual coset construction starts from a symmetry breaking pattern $G\rightarrow H$. A coset representative for $G/H$ and its Maurer-Cartan form supply desired building blocks, from which one can write down the most general local action compatible with the nonlinearly realized symmetry. In the SK extension, the fields and sources are placed on the two contour branches, rotated to $r$/$a$ basis. All the fields should be constrained by SK unitarity, positivity and, in a thermal state, dynamical KMS symmetry. This is a systematic and broadly applicable framework for constructing Goldstone effective theories, including type-A and type-B modes \cite{Akyuz:2023nbo,Hongo:2019qhi,Hongo:2024brb}. 

The main difference between our approach and the conventional coset construction resides in the input data. We do not need to introduce any particular representative of coset. Instead, we foreground globally defined geometric tensors, for instance, the Berry curvature and the metric tensor. The essential rationale behind this approach is that geometric quantities of global geometric quantities are more capable of producing the crucial Berry term needed in type-B systems covariantly. Thus the coset language is best viewed as the general symmetry and operator language, while the global formulation proposed in this work makes the physical Berry data more explicit. 

The second essential point concerns the meaning of SK doubling. Although the exact contour field space is $\mathcal{M}_1\times \mathcal{M}_2$, the physical region lies on its near-diagonal region. Thus, in our framework, we have the following fundamental geometric interpretation
\begin{equation}
\left(\pi_1, \pi_2\right) \quad \longleftrightarrow \quad \pi_r \in \mathcal{M}, \quad \pi_a \in T_{\pi_r} \mathcal{M} .
\end{equation}
That is to say, the $r$-type field is interpreted as physical Goldstone configuration and the $a$-type field is corresponding off-diagonal response/fluctuation displacement, which can not be treated as some new physical Goldstone excitation. In a given local coset chart, our formulation naturally reduces to usual $r$/$a$ basis used in SK formalism. On the contrary, in usual coset construction, the coset fields and Maurer-Cartan forms are directly doubled without any further geometric considerations.  

In summary, the constructions are complementary descriptions of the same low-energy theory. The commonly used coset method answers which SK operators are allowed by the symmetry breaking pattern and thermal constraints. The global construction formulated in this work identifies the invariant Berry data carried by a type-B sector, while the near-diagonal construction gives a precise physical and covariant meaning to its $r$/$a$ variables. The practical gain of our organization is not a new symmetry principle, but a compact way to keep Berry dynamics, dissipation, noise and dynamical KMS condition tied to the same physical phase space structure.

\section{Formal theory in practice}
\label{sec:model study}

In the preceding sections, we have presented in detail the theoretical framework for constructing the SK effective field theory suited for type-B Goldstone systems. In what follows, we apply this framework to several concrete model examples.

\subsection{Minimal model: ferromagnet}

For type-B Goldstone systems, the most well-known example is the ferromagnet. We choose the ordered state to have magnetization density $M>0$ along the third spin direction, that is,
\begin{equation}
\left[S_i, S_j\right]=i \epsilon_{i j k} S_k, \quad\left\langle S_z\right\rangle=M, \quad S U(2) \longrightarrow U(1)_{S_z} .
\end{equation}
One can easily check that the two broken generators satisfy $-i\left\langle\left[S_x, S_y\right]\right\rangle=M$. Therefore, they form one canonical pair and produce one type-B Goldstone mode, i.e., the magnon \cite{Burgess:1998ku}.

Now we apply our framework to this ferromagnetic system as a minimal example. Naturally, we need to write down all desired geometric data. Although the explicit forms of the relevant geometric quantities have been partly touched upon in the earlier section reviewing type-B Goldstone systems, we reproduce them here for the convenience of the reader and completeness of argument. The Goldstone coset manifold reads 
\begin{equation}
\mathcal{M}=SU(2)/U(1) \simeq S^2.
\end{equation}
Each point on this coset can be described by a unit vector $\boldsymbol{n}$. Conveniently, we parameterize this unit vector as
\begin{equation}
\boldsymbol{n}=(\sin \theta \cos \varphi, \sin \theta \sin \varphi, \cos \theta), \quad \boldsymbol{n}^2=1 .
\end{equation}
Based on this particular coordinate choice, the metric and Berry curvature are given by
\begin{equation}
g=\mathrm{d} \theta^2+\sin ^2 \theta \mathrm{~d} \varphi^2, \quad \Omega=M \sin \theta \mathrm{~d} \theta \wedge \mathrm{~d} \varphi.
\end{equation}
Note that in this case, the other symmetric tensor, $\mathcal{G}_{AB}$, is not an independent tensor. To clarify, one can immediately check
\begin{equation}
\mathcal{G}_{AB}=\Omega_{C A} g^{C D} \Omega_{D B}=M^2 g_{A B},
\end{equation}
which is proportional to metric tensor $g_{AB}$.

Usually, one treat Goldstone fields as fluctuation around chosen vacuum. Thus, we can introduce the following normal coordinates about the north pole of $S^2$,
\begin{eqnarray}
&& \pi^A=\theta(\cos \varphi, \sin \varphi)^A, \nonumber\\
&& \quad \theta=|\boldsymbol{\pi}|=\sqrt{\left(\pi^x\right)^2+\left(\pi^y\right)^2}, \quad A=x, y .
\end{eqnarray}
The polar angle $\theta$ can be explained as the geodesic distance in this normal coordinate plane. As a result, the essential tensors in these coordinates can be written as
\begin{eqnarray}
&& g_{A B}(\pi)  =\left(\frac{\sin \theta}{\theta}\right)^2\left(\delta_{A B}-\frac{\pi_A \pi_B}{\theta^2}\right)+\frac{\pi_A \pi_B}{\theta^2}, \nonumber\\
&& \Omega_{A B}(\pi)  =M \frac{\sin \theta}{\theta} \epsilon_{A B}, \quad \epsilon_{x y}=+1 .
\label{eq:normal coordinate}
\end{eqnarray}
It should be emphasized again that the coordinates denoted by $\pi^A$ here are applicable only within a local patch near the north pole. Tensors expressed in terms of these coordinates serve as local approximations to the globally defined tensors. Thus, in the subsequent derivation, we shall obtain the corresponding effective action directly from the global quantity $\boldsymbol{n}$, and only then carry out an expansion around the north pole.

Let $\boldsymbol{n}_r(x)$ denote the physical $r$-type field. As for the $a$-type field, we choose to represent it by a tangent vector $\boldsymbol{\xi}(x)$ satisfying $\boldsymbol{n}_r \cdot \boldsymbol{\xi}=0$. For later convenience, we also define $q\equiv |\boldsymbol{\xi}|$. Naturally, we have a unit vector $\boldsymbol{e}=\boldsymbol{\xi} / q$ for $q\neq0$. Fortunately, the short geodesic of $S^2$ connecting two contour fields is known exactly. Thus, we obtain the desired interpolation field, written as
\begin{eqnarray}
&& \boldsymbol{n}_s=\cos (s q) \boldsymbol{n}_r+\sin (s q) \boldsymbol{e}, \nonumber\\
&& \quad-\frac{1}{2} \leq s \leq \frac{1}{2}, \quad \boldsymbol{n}_{1,2}=\boldsymbol{n}_{s= \pm 1 / 2} .
\end{eqnarray}
Consequently, the exact Berry transgression is 
\begin{equation}
I_{\rm{B}}^{\mathrm{FM}}=M \int \mathrm{~d}^{d+1} x \int_{-1 / 2}^{1 / 2} \mathrm{~d} s\ \boldsymbol{n}_s \cdot\left(\partial_s \boldsymbol{n}_s \times \partial_t \boldsymbol{n}_s\right) .
\end{equation}
In this model, fortunately, the auxiliary integral of parameter $s$ can be done explicitly. After some direct algebra, one finds
\begin{equation}
\label{eq: ferromagnet Berry}
I_{\rm{B}}^{\mathrm{FM}}=\int \mathrm{d}^{d+1} x \frac{2 M \sin (q / 2)}{q}\left(\boldsymbol{n}_r \times \boldsymbol{\xi}\right) \cdot \partial_t \boldsymbol{n}_r .
\end{equation}
To proceed, we decompose the tangent vector as $\boldsymbol{\xi}=\xi^{\theta} \partial_{\theta}\boldsymbol{n}_r+\xi^{\varphi} \partial_{\varphi}\boldsymbol{n}_r$. Then 
\begin{eqnarray}
&& q^2=\left(\xi^\theta\right)^2+\sin ^2 \theta_r\left(\xi^{\varphi}\right)^2, \nonumber\\
&& \quad\left(\boldsymbol{n}_r \times \boldsymbol{\xi}\right) \cdot \dot{\boldsymbol{n}}_r=\sin \theta_r\left(\xi^\theta \dot{\varphi}_r-\xi^{\varphi} \dot{\theta}_r\right),
\end{eqnarray}
and Eq. (\ref{eq: ferromagnet Berry}) becomes a completely explicit local expression of Berry term in corresponding Lagrangian, i.e.,
\begin{equation}
\mathcal{L}_{\rm{B}}^{\mathrm{FM}}=\frac{2 M \sin (q / 2)}{q} \sin \theta_r\left(\xi^\theta \dot{\varphi}_r-\xi^{\varphi} \dot{\theta}_r\right) .
\end{equation}
 
By using Eq. (\ref{eq:normal coordinate}), we can further rewrite the above local expression of Berry term into the form suitable for near north pole region of $S^2$. This gives us
\begin{equation}
\mathcal{L}_{\rm{B}}^{\mathrm{FM}}=M \frac{2 \sin (q / 2)}{q} \frac{\sin \theta_r}{\theta_r} \epsilon_{A B} \pi_a^A \dot{\pi}_r^B,
\end{equation}
with $\theta_r=|\boldsymbol{\pi}_r|$ and $q^2=g_{AB}(\pi_r) \pi^A_a\pi^B_a$. As can be seen, the model-dependent nonlinear vertices have already been resummed. Of course, one may assume that all fields are small quantities, thereby further expanding the nonlinear action given above. 

Next, we discuss the contributions originating from conservative sector. For simplicity, we truncate free energy density at the level of second order spatial derivatives. Remember in ferromagnetic coset, we only have one rank-2 symmetric tensor. Thus, the desired form of free energy density is simply given by
\begin{equation}
\label{eq:free energy density ferro}
f_2=\frac{\rho_s}{2}\left[\left(\partial_i \theta_r\right)^2+\sin ^2 \theta_r\left(\partial_i \varphi_r\right)^2\right]=\frac{\rho_s}{2}\left(\partial_i \boldsymbol{n}_r\right)^2 .
\end{equation}
Here, the Wilsonian coefficient $\rho_s$ should be understood as spin stiffness. As a result, the corresponding covectors are directly obtained by taking functional derivative of free energy with respect to $r$-type fields, that is, 
\begin{eqnarray}
&& E_\theta\left(f_2\right)=-\rho_s\left[\nabla^2 \theta_r-\sin \theta_r \cos \theta_r\left(\partial_i \varphi_r\right)^2\right], \nonumber\\
&& E_{\varphi}\left(f_2\right)=-\rho_s \partial_i\left(\sin ^2 \theta_r \partial_i \varphi_r\right) .
\end{eqnarray}
Accordingly, the effective action of this sector is written as
\begin{eqnarray}
&& I_{\mathcal{F}}^{\rm{FM}}=\rho_s \int \mathrm{~d}^{d+1} x\Bigl\{\xi^\theta\left[\nabla^2 \theta_r-\sin \theta_r \cos \theta_r\left(\partial_i \varphi_r\right)^2\right] \nonumber\\
&& {}+\xi^{\varphi} \partial_i\left(\sin ^2 \theta_r \partial_i \varphi_r\right)\Bigr\}.
\end{eqnarray}

Finally, we construct the fluctuation and dissipation part of effective action. In Eqs. (\ref{eq:general operator}), (\ref{eq:1st order operator}) and (\ref{eq:2nd order operator}), we have already discussed possible operator classification which can be used in construction of noise and dissipation kernels. Here, some comments are needed. One may note that in Eq. (\ref{eq:general operator}), the leading term involves no additional spatial derivatives. While mathematically allowed, such kind of operators is not always physically realizable. The action terms originating from this kind of operators involve only single time derivatives, implying that they capture a local relaxation process. When spin couples to some external environment, such local relaxation makes sense. However, for the closed system under investigation in the present work, the spin degrees of freedom are conserved and do not couple to any additional dynamical degrees of freedom. As a result, the operators involving no spatial derivatives are forbidden. Hence, the minimal positive symmetric operator is $-\lambda_s D_i D_i$, with non-negative Wilsonian coefficient $\lambda_s\geq 0$. Since the geometric quantities on our coset manifold have already been given, the covariant derivative acting on an arbitrary tangent vector, denoted by $\boldsymbol{V}=(V^{\theta},V^{\varphi})$, can be written explicitly as 
\begin{eqnarray}
&& \left(D_i \boldsymbol{V}\right)^\theta  =\partial_i V^\theta-\sin \theta_r \cos \theta_r\left(\partial_i \varphi_r\right) V^{\varphi}, \nonumber\\
&& \left(D_i\boldsymbol{V}\right)^{\varphi}  =\partial_i V^{\varphi}+\cot \theta_r\left[\left(\partial_i \theta_r\right) V^{\varphi}+\left(\partial_i \varphi_r\right) V^\theta\right] .
\end{eqnarray}
Thus, the effective action of dissipation and fluctuation sector is given by
\begin{eqnarray}
&& I_{\mathrm{diss}+\mathrm{fluc}}^{\mathrm{FM}}=\int \mathrm{d}^{d+1} x  \Bigl\{-\lambda_s\Bigl[\left(D_i \boldsymbol{\xi}\right)^\theta\left(D_i \dot{\boldsymbol{n}}_r\right)^\theta \nonumber\\
&& {}+\sin ^2 \theta_r\left(D_i \boldsymbol{\xi}\right)^{\varphi}\left(D_i \dot{\boldsymbol{n}}_r\right)^{\varphi}\Bigr] \nonumber\\
&& \left.+i T \lambda_s\left[\left(\left(D_i \boldsymbol{\xi}\right)^{\theta}\right)^2+\sin ^2 \theta_r\left(\left(D_i \boldsymbol{\xi}\right)^{\varphi}\right)^2\right]\right\} .
\label{eq:diss+fluc ferro}
\end{eqnarray}
Then, the full SK effective action of ferromagnet is
\begin{equation}   
\label{eq:full ferromagnet action}
I_{\mathrm{SK}}^{\mathrm{FM}}=I_{\mathrm{B}}^{\mathrm{FM}}+I_{\mathcal{F}}^{\mathrm{FM}}+I_{\mathrm{diss}+\mathrm{fluc}}^{\mathrm{FM}}.
\end{equation}
It can be seen that the SK effective field theory constructed via our global geometric approach naturally incorporates nonlinear interactions. 

Prior to the analysis of the dispersion relation and two-point correlators, we provide a concise discussion of the DKMS transformation. If choosing north pole of coset $S^2$ as vacuum configuration, one can check that time reversal alone exchanges
the two magnetized vacua. Thus, the correct anti-unitary operation that fixes the chosen north pole vacuum can be taken as $\Theta=R_y(\pi) T$, acting by
\begin{equation}
\vartheta:(t, \boldsymbol{x}) \mapsto(-t, \boldsymbol{x}), \quad \tau:(\theta, \varphi) \mapsto(\theta,-\varphi) .
\end{equation}
It obeys desired $\tau^{\ast}g=g$ and $\tau^{\ast}\Omega=-\Omega$. The classical DKMS transformation is now written down explicitly:
\begin{eqnarray}
&& \widetilde{\theta}_r(-t, \boldsymbol{x})=\theta_r(t, \boldsymbol{x}), \nonumber\\
&& \widetilde{\varphi}_r(-t, \boldsymbol{x})=-\varphi_r(t, \boldsymbol{x}), \nonumber\\
&& \widetilde{\xi}^\theta(-t, \boldsymbol{x})=\xi^\theta(t, \boldsymbol{x})+i \beta \dot{\theta}_r(t, \boldsymbol{x}), \nonumber\\
&& \widetilde{\xi}^{\varphi}(-t, \boldsymbol{x})=-\xi^{\varphi}(t, \boldsymbol{x})-i \beta \dot{\varphi}_r(t, \boldsymbol{x}) .
\end{eqnarray}

Using the normal coordinates introduced in Eq. (\ref{eq:normal coordinate}) and keeping only quadratic terms in Eq. (\ref{eq:full ferromagnet action}), we have the following simple effective action:
\begin{eqnarray}
&& I_{\mathrm{SK}}^{(2)}=\int \mathrm{d}^{d+1} x\Bigl[M \epsilon_{A B} \pi_a^A \dot{\pi}_r^B+\rho_s \pi_a^A \nabla^2 \pi_r^A \nonumber\\
&& {}-\lambda_s \partial_i \pi_a^A \partial_i \dot{\pi}_r^A+i T \lambda_s \partial_i \pi_a^A \partial_i \pi_a^A\Bigr] .
\end{eqnarray}
After transforming into momentum space with $p=(\omega,\boldsymbol{k})$, this quadratic action is rewritten as
\begin{eqnarray}
&& I_{\mathrm{SK}}^{(2)}=-\int_p \pi_a^A(-p) \mathcal{K}_{A B}^R(p) \pi_r^B(p) \nonumber\\
&& {}+i T \lambda_s  \int_p \boldsymbol{k}^2 \pi_a^A(-p) \pi_a^A(p),
\end{eqnarray}
with corresponding kernel $\mathcal{K}_{A B}^R(p)=\left(\rho_s \boldsymbol{k}^2-i \lambda_s \boldsymbol{k}^2 \omega\right) \delta_{A B}+i M \omega \epsilon_{A B}$. To simplify the subsequent expressions, we define scalar function $\Delta_R(p)\equiv\rho_s\boldsymbol{k}^2-i\lambda_s \boldsymbol{k}^2\omega$. 

Owing to the existence of non-diagonal tensor $\epsilon$, the projection operator introduced below proves to be useful:
\begin{equation}
\label{eq:projector definition}
\mathbb{P}_\sigma=\frac{1}{2}(\delta-i \sigma \epsilon), \quad \epsilon \mathbb{P}_\sigma=i \sigma \mathbb{P}_\sigma, \quad \sigma= \pm 1.
\end{equation}
Physically, these two helicity $\sigma=\pm 1$ just correspond to positive and negative frequency branches of the same real magnon. Their dispersion relation can now be calculated by solving $\det \mathcal{K}^R=0$. Thus we have
\begin{equation}
\omega_\sigma(\boldsymbol{k})=\frac{\rho_s \boldsymbol{k}^2}{\sigma M+i\lambda_s \boldsymbol{k}^2} =\sigma \frac{\rho_s}{M} \boldsymbol{k}^2-i \frac{\rho_s \lambda_s}{M^2} \boldsymbol{k}^4+O\left(\boldsymbol{k}^6\right) .
\end{equation}
The second equality holds in small-$\boldsymbol{k}$ region. The $\boldsymbol{k}^2$ term is the ferromagnetic spin wave. Also, the dissipative $\boldsymbol{k}^4$ attenuation satisfies the spin conserving model-J scaling \cite{Hohenberg:1977ym}. 

The inverse of quadratic action gives us two-point correlation functions. The retarded correlator reads
\begin{equation}
G^{AB}_R(p)=\sum_{\sigma= \pm 1} \frac{\mathbb{P}_\sigma^{A B}}{\rho_s \boldsymbol{k}^2-\left(\sigma M+i\lambda_s \boldsymbol{k}^2\right) \omega}.
\end{equation}
Similarly, the symmetric, which is sometimes referred to as fluctuation, correlator is obtained to be
\begin{equation}
G_{\mathrm{sym}, \sigma}(p)=\frac{2 T \lambda_s \boldsymbol{k}^2}{\left(\rho_s \boldsymbol{k}^2-\sigma M \omega\right)^2+\lambda_s^2 \boldsymbol{k}^4 \omega^2},
\end{equation}
for a given helicity channel. One can thus immediately check that the classical fluctuation-dissipation theorem holds, i.e.,
\begin{equation}
    G_{\mathrm{sym},\sigma}(p)=\frac{2T}{\omega} \operatorname{Im} G_{R,\sigma}(p).
\end{equation}

As can be seen from the computed correlation functions, although the Berry term appears independently in our effective action and does not mix with other terms under the DKMS transformation, its physical information nevertheless manifests itself in the observable fluctuation and dissipation behavior of the system.

\subsection{Dissipative \texorpdfstring{$SU(2)\times U(1)$}{SU(2) × U(1)} linear sigma model}

We now turn to another interesting model, namely the dissipative $SU(2)\times U(1)$ linear sigma model. This model is invented to be an effective model describing the kaon condensation in the dense QCD matter \cite{Miransky:2001tw,Schafer:2001bq}. Before constructing its SK effective theory explicitly, we first analyze the corresponding pattern of spontaneous symmetry breaking, from which preliminary knowledge concerning the associated types of Goldstone can be extracted. Here, we choose the symmetry breaking pattern to be $G=SU(2)\times U(1)\rightarrow H=U(1)$. Given that the present discussion does not concern spin degrees of freedom, in contrast to the preceding model, the generators of $SU(2)$ are denoted by $T_i=\sigma_i/2$ for the sake of clarity. As for the additional $U(1)$, its generator is denoted by $Y=\boldsymbol{1}/2$. The corresponding algebra is thus given by
\begin{equation}
\left[T_i, T_j\right]=i \epsilon_{i j k} T_k, \quad\left[Y, T_i\right]=0 .
\end{equation}
For convenience, one choose the following decomposition as physical broken generators and unbroken generator
\begin{equation}
T_H=Y+T_3, \quad X_1=T_1, \quad X_2=T_2, \quad X_3=Y-T_3 .
\end{equation}
$T_H$ is usually interpreted as electric charge. In this basis, the algebra is rewritten as
\begin{eqnarray}
&& \left[X_1, X_2\right]=\frac{i}{2}\left(T_H-X_3\right), \nonumber\\
&& \quad\left[X_3, X_1\right]=-i X_2, \nonumber\\
&& \quad\left[X_3, X_2\right]=i X_1.
\end{eqnarray}

For a homogeneous finite density ground state, let $n_H\equiv\langle Q_{T_H}\rangle$ and $n_a\equiv \langle Q_{X_a}\rangle$, where $Q_i$ are the conserved charges associated with corresponding generators. For the unbroken generators, we naturally have $n_H=0$. In the broken subspace, the residual basis freedom can be used to align the charge density vector as $\left(n_1, n_2, n_3\right)=\left(0,0, n_0\right)$, with $n_0>0$. Accordingly, the desired density matrix is given by
\begin{equation}
\left(\rho_{a b}\right)=\left(\begin{array}{ccc}
0 & \rho_B & 0 \\
-\rho_B & 0 & 0 \\
0 & 0 & 0
\end{array}\right), \quad \quad \rho_B \equiv-\frac{n_0}{2} .
\end{equation}
In this case, $\rho$ is degenerate. It follows immediately that
\begin{eqnarray}
&& N_B=\frac{1}{2} \operatorname{rank} \rho=1, \nonumber\\
&& \quad N_A=N_{\mathrm{BS}}-\operatorname{rank} \rho=1, \nonumber\\
&& \quad N_{\mathrm{NG}}=N_A+N_B=2 .
\end{eqnarray}
The pair $(X_1,X_2)$ is therefore the type-B sector, while $X_3$ supplies the type-A Goldstone. Compared with the ferromagnet example discussed earlier, the present model is somewhat more involved, as it exhibits a mixing of different types of Goldstone modes. However, in this work we discuss these two different types of Goldstone modes separately for simplicity and postpone a thorough discussion to future work. 

We now evaluate desired geometric quantities for this linear sigma model. From our chosen symmetry breaking pattern, we have the coset
\begin{equation}
\mathcal{M}=\frac{S U(2) \times U(1)}{U(1)} \simeq S^3.
\end{equation}
It has already been pointed out that $S^3$ is a $U(1)$ bundle over $S^2$ \cite{Nakahara:2003}. This interesting geometric structure greatly simplifies our construction of SK effective action. As will be seen in the subsequent derivation, this geometric structure renders the effective theory of the type-B sector in the present model almost identical to that of the ferromagnet system discussed earlier, with the only difference residing in the physical interpretation of certain Wilsonian coefficients. 

A commonly used parameterization of this coset is given by a unit complex doublet, i.e.,
\begin{equation}
\label{eq:complex parameterization}
Z=\binom{e^{i(\psi+\varphi) / 2} \sin (\theta / 2)}{e^{i(\psi-\varphi) / 2} \cos (\theta / 2)} .
\end{equation}
One can see that, in this parameterization, $(\theta,\varphi)$ describe the degrees of freedom on base space $S^2$ and $\psi$ corresponds to $U(1)$ bundle structure. Based on this parameterization, we thus have the metric of such coset, which is given by
\begin{eqnarray}
&& \mathrm{d}s^2=v^2 \mathrm{d} Z^{\dagger} \mathrm{d} Z \nonumber\\
&& {}=\frac{v^2}{4}\left[\mathrm{d} \theta^2+\sin ^2 \theta \mathrm{d} \varphi^2+(\mathrm{d} \psi-\cos \theta \mathrm{d} \varphi)^2\right],
\end{eqnarray}
where $v$ denotes the symmetry breaking scale that normalizes the sigma model metric. Further, the Berry connection can be also obtained by this parameterization,
\begin{equation}
\alpha=-i Z^{\dagger} \mathrm{d} Z=\frac{1}{2}(\mathrm{d} \psi-\cos \theta \mathrm{d} \varphi) .
\end{equation}
According to the previous discussion of $\rho_{ab}$, we know that the Berry curvature here should proportional to $n_0$. For this linear sigma model, we choose the matching condition to be $n_0=2\mu v^2$. This condition physically makes sense. Roughly speaking, $v^2$ describes the condensate density, while the chemical potential $\mu$ itself reflects the basic physics that the charge response originates from a finite density background. Accordingly, our convention of Berry curvature reads
\begin{equation}
\Omega=\mathrm{d}\left(2 \mu v^2 \alpha\right)=\mu v^2 \sin \theta \mathrm{~d} \theta \wedge \mathrm{d} \varphi.
\end{equation}
One can see that the $\psi$ field does not contribute to Berry curvature, thus corresponds to type-A sector. On the contrary, $(\theta,\varphi)$ form type-B sector. Similar to ferromagnet, one can also use $\boldsymbol{n}_r(x)\in S^2$ to be the physical field and let $\boldsymbol{\xi}(x)\in T_{\boldsymbol{n}_r}S^2$ be its a-type tangent field. Consequently, the effective action of type-B sector of this linear sigma model can be immediately written down. 

Substituting $M\rightarrow \mu v^2$ in the Eq. (\ref{eq: ferromagnet Berry}) yields the Berry term for this sigma model. Also, substituting $\rho_s\rightarrow v^2/2$ in the Eq. (\ref{eq:free energy density ferro}) gives us the conservative sector for this model. A few additional remarks on the fluctuation and dissipation sectors are in order here. Remember that the symmetric operator classified in our general EFT has the form $\mathbb{O}=V-D_i M^{i j} D_j+D_i D_j Q^{i j ; k l} D_k D_l$. The ferromagnet example already illustrated that the specific choice of the leading order operator is dictated by the physical demands of the problem at hand. In order to compare with some previous work, for example \cite{Minami:2015nki}, we now allow the system to undergo local relaxation, unlike in the ferromagnet case, where spin degrees of freedom are required to be conserved. Thus, the leading contribution comes from $V$ part. Because tensors $P^{(1)}$ and $P^{(2)}$ are proportional to each other in the present model, we only have the following leading order terms
\begin{equation}
I^{\mathrm{LS}}_{\mathrm{diss}+\mathrm{fluc}}=-\frac{\gamma_0 v^2}{2} \int \mathrm{d}^{d+1}x \  \boldsymbol{\xi} \cdot \dot{\boldsymbol{n}}_r+\frac{i T \gamma_0 v^2}{2} \int \mathrm{d}^{d+1} x\  \boldsymbol{\xi}^2,
\end{equation}
 with $\gamma_0\geq 0$ measuring local relaxation rate of the system. Hence, we arrive at the SK effective field theory action for the type-B Goldstone sector in this dissipative linear sigma model.

Comparing the ferromagnet with the dissipative linear sigma model, one sees that, as long as the geometry associated with the type-B sector is the same, which in both cases is $S^2$, the operators appearing in the corresponding SK effective action are also identical. The only difference resides in the physical interpretation of the Wilsonian coefficients multiplying these operators, which depends on the specific system under consideration. 

In order to further calculate the dispersion relation and two-point correlation functions, we expand the effective action around the north pole and truncate at quadratic order. To clarify, the local coordinate near the north pole can be given by
\begin{equation}
z=\mathrm{e}^{i \varphi} \tan \frac{\theta}{2}, \quad \pi^1+i \pi^2=\sqrt{2} z.
\end{equation}
Thus, the quadratic effective action of type-B sector reads
\begin{eqnarray}
&& I^{B,(2)}_{\mathrm{SK}}=v^2 \int \mathrm{d}^{d+1}x \Bigl[-\partial_i \pi_a^A \partial_i \pi_r^A-\gamma_0 \pi_a^A \dot{\pi}_r^A \nonumber\\
&& {}+2\mu \epsilon_{A B} \pi_a^A \dot{\pi}_r^B+i T \gamma_0 \pi_a^A \pi_a^A\Bigr],
\end{eqnarray}
where the superscript "$B$" on the left hand side denotes the type-B sector. Similarly, the superscript "$A$" will be used to refer to the type-A sector. 

At quadratic order, the type-A and type-B sectors are independent of each other, and it is therefore reasonable to treat their effective actions separately. It must be emphasized, however, that once one goes beyond the quadratic order, mixing between the two sectors is to be expected in a proper construction of the effective theory, which would render the theory considerably more involved. Such a more general and more complicated situation lies beyond the scope of the present work.

To obtain the effective action, firstly we need to identify correct field description of type-A sector. Since only the quadratic order effective action needs to be discussed at this stage, we may proceed by investigating directly around the north pole. Set $\theta=0$ in our parameterization, Eq. (\ref{eq:complex parameterization}), one get
\begin{equation}
Z_{\mathrm{N}}=\binom{0}{e^{i \chi}}=e^{i \chi X_3}\binom{0}{1},
\end{equation}
with $\chi \equiv (\psi-\varphi)/2$. Thus the correct degree of freedom along the broken generator $X_3=Y-T_3$ which should correspond to type-A sector is $\chi$ field. 

The leading SK effective action for type-A Goldstone system has been discussed extensively in previous work \cite{Akyuz:2023nbo,Hongo:2019qhi}. Since the primary goal of the present work is to discuss the construction of the SK effective field theory for type-B Goldstone modes, we refrain from elaborating further on the type-A Goldstone sector. Interested readers are referred to the aforementioned literature. Consequently, we only quote the final result. Matching the corresponding low-energy constants to the present linear sigma model yields
\begin{equation}
I_{\mathrm{SK}}^{A,(2)}=2 v^2 \int \mathrm{d}^{d+1}x\left[-\partial_i \chi_a \partial_i \chi_r-\gamma_1 \chi_a \dot{\chi}_r+i T \gamma_1 \chi_a^2\right],
\end{equation}
with $\gamma_1\geq 0$. The factor of "2" in front of the action originates from the definitional convention for the field variables, and can of course be eliminated by a field redefinition.

Once the corresponding effective action at quadratic order has been obtained, the computation of the dispersion relation and the two-point correlation functions follows naturally. For type-B sector, the projector defined in Eq. (\ref{eq:projector definition}) remains useful. After some direct algebra, the retarded correlator of type-B sector is given by 
\begin{equation}
G_R^B(\omega,\boldsymbol{k})=\frac{1}{v^2} \sum_{\sigma= \pm 1} \frac{\mathbb{P}_\sigma}{\kappa_\sigma^R}, \quad \kappa_\sigma^R=\boldsymbol{k}^2-\left(2\sigma\mu +i \gamma_0\right) \omega .
\end{equation}
Similarly, the retarded correlator of type-A sector is calculated to be 
\begin{equation}
G_R^A(\omega, \boldsymbol{k})=\frac{1}{2 v^2 \kappa_{\chi}^R} ,\quad \kappa_{\chi}^R=\boldsymbol{k}^2-i\gamma_1 \omega.
\end{equation}

It is a easy task to extract physical excitation from the pole structure of retarded correlator. Thus, for type-B Goldstone, we have the following dispersion relation
\begin{equation}
\omega_\sigma^B(\boldsymbol{k})=\frac{\boldsymbol{k}^2}{2\sigma\mu +i \gamma_0}=\frac{2\sigma\mu-i \gamma_0}{(2\mu)^2+\gamma_0^2} \boldsymbol{k}^2.
\end{equation}
As can be seen, the type-B Goldstone in dissipative system is still a propagating mode. Besides, dissipation occurs at $\mathcal{O}(\boldsymbol{k}^2)$ when local relaxation is allowed, in contrast to the $\mathcal{O}(\boldsymbol{k}^4)$ behavior found in the conserved case. To avoid any potential misunderstanding, we emphasize once more that the two values of $\sigma$ here do not refer to two independent type-B Goldstone modes. Rather, they correspond to the positive and negative frequency branches of a single type-B Goldstone mode.

For comparison, the type-A Goldstone mode in a dissipative system exhibits the following physical excitation:
\begin{equation}
\omega^A(k)=-\frac{i}{\gamma_1}\boldsymbol{k}^2 ,
\end{equation}
which is a purely diffusive mode. Therefore, once dissipative effects are introduced, type-A and type-B Goldstone modes exhibit completely distinct behaviors. It should be emphasized that the present result is consistent with that obtained in \cite{Minami:2015nki}, where the same model was analyzed using a different approach. 

\section{Summary and discussion}
\label{sec:summary}

Owing to its connection with the Berry curvature, it is natural to study type-B Goldstone modes from a geometric perspective. This work demonstrates that, when one attempts to formulate the effective theory of type-B Goldstone systems further based on the SK contour, certain intriguing geometric structures reveal themselves. For systems at finite temperature, after taking the classical limit, the region of physical interest is not the one obtained by simply doubling the coset manifold, but rather a submanifold thereof, referred to in this work as the near-diagonal region. In this proposed description, the $r$-type field is identified as the physical Goldstone field and the $a$-type field corresponds to its tangent vector. 

Most interestingly, we integrate the Berry curvature over the strip region joining the two SK time contour, which is called transgression construction in this work, and this procedure gives us an exact term in effective action without choosing a coordinate dependent local potential. The near-diagonal expansion of this exact transgression only contains odd powers of $a$-type field. The leading term is just the standard type-B kinetic coupling, while higher nonlinear interaction terms are totally fixed by Berry curvature and introduce no new Wilsonian coefficients. For system at finite temperature, the DKMS invariance is an essential physical constraint. Our construction of Berry term is compatible with DKMS invariance in the classical limit up to some boundary contributions. 

The remaining sectors of our SK effective theory were organized by globally invariant tensors on the manifold under investigation. Based on this point, we further discuss the classification of possible operators. By using constraints originating from SK formalism, we have also obtained the basic conditions satisfied by the kernels corresponding to the different sectors. These constraints, together with the physical relations they entail, assist us in writing down the correct SK effective field theory for type-B Goldstone modes. 

After finishing the discussion of rather formal construction of effective theory, we investigate two concrete examples with different symmetry structures. The ferromagnet is the simplest system that contains only type-B Goldstone modes, and can therefore serve as a clean testing ground for the construction scheme developed in this work. By comparison, the dissipative linear sigma model is somewhat more involved, as it contains both a type-B sector and a type-A sector. Thus, this model provides us with a direct comparison between the behavior of type-A and type-B Goldstone modes in the presence of dissipative effects. In summary, the type-B Goldstone mode remains propagating, albeit with a finite lifetime induced by dissipative effects. By contrast, the type-A Goldstone mode is purely diffusive, its frequency being purely imaginary. The conclusions drawn from the study of these two models within the framework of this work are consistent with existing results in the literature \cite{Minami:2015nki,Hongo:2019qhi}. 

With regard to this work, there are two limitations that merit being mentioned. First, the local operator analysis uses only the classical limit of DKMS. The full quantum transformation contains a finite thermal translation and an infinite derivative series. Thus, its interplay with a grand canonical thermal twist, nonlinear response fields and external sources remains to be established. Second, higher order contributions have not been treated in comparable detail.  Although our transgression construction fixes a series expansion in the near-diagonal region and the kernel notation used in this work admits higher derivative contributions, quantum DKMS constraints on possible non-Gaussian terms are absent. Also, the examples discussed in this work were restricted to quadratic effective actions, leaving nonlinear type-A/type-B mixing and loop effects outside the present results.

Based on the above shortcomings, the issues that need to be further discussed and investigated in the future become quite clear. On the one hand, it is necessary to discuss the quantum DKMS condition more systematically and in greater depth, and to genuinely incorporate it into the construction of effective theories, with the expectation that it will impose more stringent constraints on the nonlinear interaction terms. It should be mentioned that the implementation of the DKMS symmetry at all orders in $\hbar\omega/T$ was recently discussed in \cite{Firat:2025upx}. On the other hand, we may extend our covariant operator classification to higher-order terms, and further discuss the possible mixing between type-A and type-B Goldstone modes that can arise at nonlinear order, as well as the impact of non-Gaussian noise on the behavior of the system. We anticipate that these further investigations will test the framework developed in this work and deepen our understanding of the physical properties of dissipative systems hosting Goldstone excitations.

\begin{acknowledgments}

We thank Zi-Hao Liu for helpful discussions and sharing many insights with us. This work is supported by the National Natural Science Foundation of China (NSFC) Grant Nos. 12235016, 12221005. 

\end{acknowledgments}

\appendix

\section{Derivation of Eq. (\ref{eq:key relation})}
\label{app:A}

Although basic mathematics behind Eq. (\ref{eq:key relation}) can be explained by the so-called coadjoint orbit geometry, it is beyond the scope of this work. Interested readers can consult some classical materials, for example \cite{Delacretaz:2022ocm}. Instead, we can derive such relation in a more physical way, directly from the Goldstone states.

Let the vacuum be $\ket{0}$. A slowly varying Goldstone modes can be locally parameterized by a family of states
\begin{equation}
|\pi\rangle=U(\pi)|0\rangle, \quad U(\pi)=\exp \left(i \pi^a Q_a\right),
\end{equation}
with $Q_a$ denoting the broken generators. For clarity, the spatial dependence of such states is suppressed. The Berry connection on this family of states is 
\begin{equation}
A_a(\pi)=-i\langle\pi| \partial_a|\pi\rangle=-i\langle 0| U^{-1} \partial_a U|0\rangle .
\end{equation}
The corresponding Berry curvature is
\begin{equation}
\Omega_{a b}=\partial_a A_b-\partial_b A_a.
\end{equation}
Because the final expectation value will be evaluated with respect to vacuum state, we only need to calculate Berry curvature at $\pi=0$. Accordingly, we have
\begin{equation}
\left.\partial_a|\pi\rangle\right|_ 0=i Q_a|0\rangle .
\end{equation}
By using the standard identity for Berry curvature,
\begin{equation}
\Omega_{a b}=-i\left(\left\langle\partial_a \pi \mid \partial_b \pi\right\rangle-\left\langle\partial_b \pi \mid \partial_a \pi\right\rangle\right),
\end{equation}
we obtain
\begin{eqnarray}
&& \Omega_{a b}(0)  =-i\left[\langle 0|\left(-i Q_a\right)\left(i Q_b\right)|0\rangle-\langle 0|\left(-i Q_b\right)\left(i Q_a\right)|0\rangle\right] \nonumber\\
&& {}=-i\langle 0| Q_a Q_b-Q_b Q_a|0\rangle  =-i\left\langle\left[Q_a, Q_b\right]\right\rangle \nonumber\\
&& {}=\rho_{ab},
\end{eqnarray}
which is the desired Eq. (\ref{eq:key relation}). 

One comment can be made here. From the basic geometric interpretation, Berry curvature measures the geometric phase accumulated by making an infinitesimal loop in
parameter space. Consider a small parallelogram generated by first shifting in direction $a$, then in direction $b$, then shifting in the opposite ordering. In group language this infinitesimal loop is controlled by the group commutator
\begin{equation}
e^{-i \epsilon Q_b} e^{-i \epsilon Q_a} e^{i \epsilon Q_b} e^{i \epsilon Q_a}=\exp \left(-\epsilon^2\left[i Q_a, i Q_b\right]+O\left(\epsilon^3\right)\right) .
\end{equation}
Taking the vacuum expectation value gives precisely the Berry curvature at the origin point ($\pi=0$) of the coset manifold. This is the intuitive content of 
\begin{equation}
    \Omega_{ab}(0)\sim \braket{[Q_a,Q_b]}.
\end{equation}
In general, if we go away from the origin, the Berry curvature will receive corrections which are depend of Goldstone fields $\pi$.

\section{A proof of DKMS invariance of Berry transgression}
\label{app:B}

In this Appendix, the covariant Berry transgression is kept
unexpanded in the SK difference field. The homogeneous
anti-unitary operation of the DKMS transformation has already been accounted for in Eq. (\ref{eq:homo constraint}). For ease of reference, we reproduce this constraint here once more: if
\begin{equation}
  \label{eq:exact section parity}
  \epsilon_{\rm B}\tau^\ast\Omega=\Omega ,
\end{equation}
then the DKMS image of the Berry transgression reduces to the same transgression evaluated on the shifted tangent vector. Thus
\begin{equation}
  \label{eq:exact section shift only}
  \widetilde I_{ B}-I_{B}
  =I_{B}[\xi+i K_\beta]-I_{B}[\xi],
\end{equation}
where the possible coset space rotation has been absorbed into the condition given by Eq. (\ref{eq:exact section parity}). The only remaining task is to show that the right hand side of Eq. (\ref{eq:exact section shift only}) is a total derivative.

Now we introduce a real interpolation parameter $\alpha$ and set
\begin{eqnarray}
&& \xi_\alpha=\xi+\alpha K_\beta, \nonumber\\
&& \quad
  \Pi_\alpha(s,x)
  =
  \exp_{\pi_r(x)}\!\left(s\,\xi_\alpha(x)\right), \nonumber\\
&& \  \text{with} \ 
  -\frac{1}{2}\leq s\leq\frac{1}{2}.
\end{eqnarray}
For real $\alpha$, we have the following analogous transgression integral:
\begin{equation}
  I_{ B}[\xi_\alpha]
  =
  \int_Xd^{d+1}x
  \int_{-1/2}^{1/2}d s\,
  \Omega(Q_s,Q_t),
\end{equation}
with some new definitions
\begin{equation}
    Q_s=\partial_s \Pi_\alpha, \quad Q_\alpha=\partial_\alpha \Pi_\alpha, \quad Q_t=D_t \Pi_\alpha .
\end{equation}

Pulling back the closed two form field $\Omega$ to the new parameter space $(s,\alpha,t)$ gives the following identity:
\begin{equation}
  0=\partial_\alpha\Omega(Q_s,Q_t)
  -\partial_s\Omega(Q_\alpha,Q_t)
  +D_t\Omega(Q_\alpha,Q_s).
\end{equation}
Hence we immediately have
\begin{equation}
  \partial_\alpha\Omega(Q_s,Q_t)
  =
  \partial_s\Omega(Q_\alpha,Q_t)
  -
  D_t\Omega(Q_\alpha,Q_s).
\end{equation}
After integration over the transgression interval, the key equation manifests itself:
\begin{eqnarray}
&& \partial_\alpha I_{B}\left[\xi_\alpha\right]=\int_X d^{d+1} x\Bigl[\left.\Omega\left(Q_\alpha, Q_t\right)\right|_{s=-1 / 2} ^{s=1 / 2} \nonumber\\
&& {}-D_t \int_{-1 / 2}^{1 / 2} d s \,\Omega\left(Q_\alpha, Q_s\right)\Bigr] .
\label{eq:master KMS identity}
\end{eqnarray}
It should be emphasized that Eq. (\ref{eq:master KMS identity}) is the  transgression identity without any local expansion.

Obviously, the second term in Eq. (\ref{eq:master KMS identity}) is a total derivative and does not contribute. Consequently, the remaining part is the endpoint term. To simplify expressions, we further define
\begin{equation}
  q_\pm(\alpha,x)=\Pi_\alpha(\pm1/2,x),
  \quad
  Y_\pm(\alpha,x)=\left.Q_\alpha\right|_{s=\pm1/2}.
\end{equation}
Denote the local patch containing the endpoint curves as $\mathcal{U}$. Now we need to assume that the following condition holds,
\begin{equation}
  \label{eq:exact section endpoint Hamiltonian}
  \left.\iota_{Y_\pm(\alpha)}\Omega\right|_{\mathcal U}
  =d h_\pm(\alpha),
\end{equation}
for local scalar functions $h_\pm(\alpha)$ defined on $\mathcal{U}$. This assumption has already been used when discussing DKMS invariance of leading-order term in expansion of transgression. Also, similar to the discussion about leading-term made in main text, a global obstruction would be the corresponding de Rham cohomology $H^1(\mathcal U)$.

When Eq. (\ref{eq:exact section endpoint Hamiltonian}) holds, we have
\begin{equation}
  \left.\Omega\left(Q_\alpha, Q_t\right)\right|_{s= \pm 1 / 2}=\mathrm{d} h_{ \pm}(\alpha)\left(D_t q_{ \pm}\right)=D_t h_{ \pm}(\alpha) .
\end{equation}
Substituting this into Eq. (\ref{eq:master KMS identity}) gives
\begin{equation}
  \partial_\alpha I_{ B}[\xi_\alpha]=\int_Xd^{d+1}x\,D_t b_\alpha ,
\end{equation}
with
\begin{equation}
  b_\alpha=h_{+}(\alpha)-h_{-}(\alpha)-\int_{-1 / 2}^{1 / 2} d s\, \Omega\left(Q_\alpha, Q_s\right) .
\end{equation}
Next, we can directly integrate over $\alpha$ and this procedure yields
\begin{equation}
    \label{eq:difference identity}
    I_{B}\left[\xi_\alpha\right]-I_{B}[\xi]=\int_X d^{d+1} x D_t \mathcal{B}_{\mathrm{ex}}(\alpha),
\end{equation}
with
\begin{eqnarray}
&& \mathcal{B}_{\mathrm{ex}}(\alpha)=\int_0^\alpha d a\Bigl[h_{+}(a)-h_{-}(a) \nonumber\\
&& {}-\int_{-1 / 2}^{1 / 2} d s \,\Omega\left(Q_a, Q_s\right)\Bigr].
\end{eqnarray}
In the near-diagonal region, which is a local patch of coset manifold, the right hand side of Eq. (\ref{eq:difference identity}) is analytic in $\alpha$, so the DKMS shift can be obtained by evaluating the same expression at $\alpha=i$:
\begin{equation}
  \label{eq:exact KMS boundary}
  \widetilde{I}_{B}-I_{B}=\int_X d^{d+1} x D_t \mathcal{B}_{\mathrm{ex}}(\mathrm{i}).
\end{equation}

Combining Eq. (\ref{eq:exact section shift only}) and Eq. (\ref{eq:exact KMS boundary}), the full DKMS variation of the Berry transgression is a desired total derivative:
\begin{equation}
  \widetilde I_{ B}-I_{ B}=\int_X d^{d+1}x\,D_t\mathcal B_{\rm ex}(i).
\end{equation}
Thus we show that the covariant Berry transgression is DKMS invariant up to a boundary term, provided the anti-unitary condition Eq. (\ref{eq:exact section parity}) and the local assumption Eq. (\ref{eq:exact section endpoint Hamiltonian}) hold.

\bibliography{ref}

\end{document}